\documentclass[twocolumn,superscriptaddress,amsfont,amssymb,amsmath, showpacs,balancelastpage, nofootinbib]{revtex4-1}
\usepackage{CJKutf8}

\usepackage{amsthm}

\usepackage[utf8]{inputenc}
\usepackage{amsmath,amssymb,amsfonts,stmaryrd}
\usepackage{physics}
\usepackage{dsfont} 
\usepackage{graphicx}
\usepackage{xcolor}
\usepackage{graphicx}
\usepackage{cancel}
\usepackage{natbib}
\usepackage{color}
\usepackage{tikz}
\usepackage{mathrsfs}
\usepackage{relsize}
\usepackage{multirow}
\usepackage[linktocpage]{hyperref}
\usepackage[all]{xy}
\hypersetup{colorlinks=true,citecolor=blue,linkcolor=blue, urlcolor=blue, breaklinks=true}

\usepackage[mathscr]{euscript}
\usepackage[scr=boondox]{mathalpha}

\usepackage{float}

\usepackage{bm} 
\usepackage{subfigure} 
\usepackage{environ}
\usepackage{url}
\usepackage[margin=0.75in]{geometry}

\usepackage{mathtools}

\newcommand{\OO}{{\text{o}}}

\newcommand{\Z}{{\mathbb Z}}

\def\U{U(1)}

\NewEnviron{eqs}{%
\begin{equation}\begin{split}
    \BODY
\end{split}\end{equation}}

\definecolor{dkgreen}{rgb}{0,0.5,0}

\theoremstyle{definition}

\theoremstyle{remark}

\begin{document}

\begin{CJK*}{UTF8}{bsmi}

\title{Crystalline invariants of fractional Chern insulators}

\author{Ryohei Kobayashi}
\affiliation{Department of Physics, Joint Quantum Institute, and Condensed Matter Theory Center, University of Maryland, College Park, Maryland 20742, USA}
\author{Yuxuan Zhang}
\affiliation{Department of Physics, Joint Quantum Institute, and Condensed Matter Theory Center, University of Maryland, College Park, Maryland 20742, USA}
\author{Naren Manjunath}
\affiliation{Perimeter Institute for Theoretical Physics, Waterloo, Ontario N2L 2Y5, Canada}
\author{Maissam Barkeshli}
\affiliation{Department of Physics, Joint Quantum Institute, and Condensed Matter Theory Center, University of Maryland, College Park, Maryland 20742, USA}

\begin{abstract}
In the presence of crystalline symmetry, topologically ordered states can acquire a host of \it symmetry-protected \rm invariants. These determine the patterns of crystalline symmetry fractionalization of the anyons in addition to fractionally quantized responses to lattice defects. Here we show how ground state expectation values of partial rotations centered at high symmetry points can be used to extract crystalline invariants. Using methods from conformal field theory and G-crossed braided tensor categories, we develop a theory of invariants obtained from partial rotations, which apply to both Abelian and non-Abelian topological orders. We then perform numerical Monte Carlo calculations for projected parton wave functions of fractional Chern insulators, demonstrating remarkable agreement between theory and numerics. For the topological orders we consider, we show that the Hall conductivity, filling fraction, and partial rotation invariants fully characterize the crystalline invariants of the system. Our results also yield invariants of continuum fractional quantum Hall states protected by spatial rotational symmetry. 

\end{abstract}

\maketitle

\end{CJK*}


A fundamental question in condensed matter physics is to understand the role of crystalline symmetry in distinguishing phases of matter. In the context of topological phases of matter, crystalline symmetry can significantly expand the set of distinct possible phases \cite{wen2002quantum,wen04,hasan2010,barkeshli2012a,Essin2013SF,Essin2014spect,barkeshli2019,Benalcazar2014,ando2015,Chiu2016review,hermele2016,barkeshli2019tr,zaletel2017,song2017,Huang2017,Shiozaki2017point,Kruthoff2017TCI,Bradlyn2017tqc,Thorngren2018,Liu2019ShiftIns,manjunath2021cgt,manjunath2020FQH,manjunath2022mzm,herzogarbeitman2022interacting,zhang2022fractional,zhang2023complete,manjunath2023classif,sachdev2023quantum,kobayashi2024}. For example, discrete translation and rotation symmetries might permute anyons, causing lattice defects to become non-Abelian \cite{barkeshli2012a,barkeshli2019}. Anyons can themselves carry fractional quantum numbers under the crystalline symmetry, giving rise to crystalline symmetry fractionalization \cite{Jalabert1991,cheng2016lsm,Sachdev_2018,sachdev1999translational,Essin2013SF,Essin2014spect,qi2015detecting,zaletel2017,barkeshli2019tr,manjunath2021cgt,manjunath2020FQH}. Finally, the system can have fractional quantized responses to lattice defects \cite{van2018dislocation,Liu2019ShiftIns,Li2020disc,manjunath2021cgt,manjunath2020FQH,zhang2022fractional,zhang2022pol}, in analogy to fractional quantized Hall conductivity. Despite significant progress over the last several decades, there are still important open questions about how to define and extract topological invariants that arise due to crystalline symmetry, particularly in fractionalized topologically ordered phases with anyons. This question has gained renewed urgency, given the experimental discovery of fractional Chern insulators (FCIs) \cite{kol1993fractional,hafezi2007,tang2011high,regnault2011,sun2011,parameswaran2013,neupert2011fractional} in crystalline two-dimensional materials  \cite{Spanton2018FCI,cai2023signatures,lu2024fractional} and ultracold atoms \cite{semeghini2021probing}.

In this paper we develop a theory of how to extract many-body topological invariants protected by crystalline symmetry from ground state wave functions of topologically ordered systems. The results presented here apply to fractional Chern insulators (FCIs) of bosons and quantum spin liquids with crystalline symmetry, and also have implications for fractional quantum Hall states with continuous spatial symmetry. In particular, we develop a theory of ground state expectation values of \it partial rotations \rm centered at high symmetry points for systems with intrinsic topological order. We show how such partial rotation expectation values can be used to completely characterize the crystalline invariants, and in particular determine the crystalline symmetry fractionalization and defect responses. This significantly extends recent work \cite{zhang2022fractional,zhang2023complete,manjunath2023classif} that shows how partial rotations can be used to completely characterize invertible topological phases \cite{barkeshli2021invertible,aasen2021characterization}, which have no anyons. It also adds to a line of work showing how topological invariants are encoded in bulk ground state wave functions \cite{levin2006,kitaev2006topological,shiozaki2017invt,dehghani2021,cian2021,cian2022extracting,Kim2022ccc,fan2022,Kobayashi2024hcc,zhang2023complete,kobayashi2024, kobayashi2024hh}.

We focus on systems with $U(1)$ charge conservation symmetry and orientation-preserving wallpaper group symmetries, corresponding to the symmetry group $G = U(1) \times_\phi [\mathbb{Z}^2 \rtimes \mathbb{Z}_M]$, for $M = 1,2,3,4,6$, where $\phi$ denotes a rational magnetic flux $\phi = 2\pi p /q$ per unit cell. Topological orders with such symmetries were systematically characterized and classified in \cite{manjunath2021cgt,manjunath2020FQH} by utilizing the theoretical framework of G-crossed braided tensor categories (BTCs) for symmetry-enriched topological (SET) phases \cite{barkeshli2019}. The results of this paper show how the invariants obtained in  \cite{manjunath2021cgt,manjunath2020FQH} can be extracted from partial rotations.

We further use the parton construction 
\cite{baskaran1988gauge,jain1989incompressible,wen1991prl,wen1992theory,wen1999projective,lee2006doping,wen2002quantum,barkeshli2010effective,mcgreevy2011fractional,wen04,sachdev2023quantum} to obtain trial wave functions for crystalline symmetry-enriched FCIs by projecting wave functions of free fermion Chern insulators. Using the parton effective field theory, we predict the crystalline invariants of the projected FCI wave functions in terms of the crystalline invariants of the parton free fermion states. Utilizing large-scale numerical Monte Carlo calculations, we then compare the numerical results of the partial rotation expectation values to the theoretical predictions, finding remarkable agreement. 

Most of the main text is focused on discussing at length a paradigmatic example, the $1/2$-Laughlin topological order on the square lattice. We provide theoretical predictions for partial rotations for general bosonic topological orders in Sec.~\ref{sec:generalboson}. 

\section{$1/2$-Laughlin on square lattice}\label{sec:laughlin}
The $1/2$-Laughlin topological order is described by $U(1)_2$ Chern-Simons (CS) theory, and arises in the description of bosonic fractional quantum Hall states, chiral spin liquids \cite{kalmeyer1987,wen1989csl,bauer2014chiral}, and FCIs. The topological order consists of two topologically distinct charges, $\{[0], [1]\}$, where $[1]$ labels the semion with topological twist $\theta_{[1]} = e^{i\pi/2}$. We have the fusion rules $[a] \times [b] = [a + b]$, where the brackets imply mod $2$ reduction.  

\subsection{$G = U(1) \times \mathbb{Z}_4$}

For now let us assume the global symmetry is $G = \U \times \mathbb{Z}_4$, where $\mathbb{Z}_4$ is a $4$-fold spatial rotational symmetry (also denoted $C_4$). We will include lattice translational symmetries afterwards. Following \cite{manjunath2021cgt}, the symmetry fractionalization class can be specified by two anyons, $[v], [s] \in \{[0], [1]\}$. $v$ is the charge vector (vison) while $s$ is a crystalline analog of the spin vector \cite{Wen1992shift}. $v$ determines Hall conductivity via $\sigma_H = \frac{1}{2\pi} (\frac{v^2}{2} + 2 k_1)$ in natural units. $v$ also determines the fractional charge of the anyon $a\in\{[0],[1]\}$, $Q_a = a v/2 \mod 1$. $s$ determines the discrete shift, which specifies the charge bound to lattice disclinations and fractional orbital angular momentum of the anyons, $L_a = a s /2 \mod 1$. The topological quantum numbers are further specified by three integers $k_1, k_2, k_3 \in \mathbb{Z}$. These invariants appear as coefficients in an effective CS theory:
\begin{align}
\label{effAc}
    \mathcal{L} = \frac{2}{4\pi} a d  a - \frac{v}{2\pi} A d  a - \frac{s}{2\pi} \omega d  a + \mathcal{L}_{\text{SPT}},
\end{align}
where $\mathcal{L}_{\text{SPT}} =  -(\frac{k_1}{2\pi} A d  A  + \frac{k_2}{2\pi} A d  \omega +  \frac{k_3}{2\pi} \omega d  \omega)$.~\footnote{We note that the Hall conductivity appears in the response action as $-\frac{\sigma_H}{4\pi}AdA$ in our convention, which leads to the formula $\sigma_H = \frac{1}{2\pi} (\frac{v^2}{2} + 2 k_1)$.}
Here $a$ is a dynamical $U(1)$ gauge field, $A$ is the background $U(1)$ gauge field, and $\omega$ is a background $\mathbb{Z}_4$ gauge field of the crystalline symmetry. Mathematically $\omega$ is treated as a real-valued 1-form with quantized holonomies $\oint \omega \in \frac{2\pi}{4} \mathbb{Z}$. Note we have  equivalences $(s,k_2, k_3) \sim (s - 2, k_2+v, k_3+s - 1 )$ and $(v, k_1,k_2) \sim (v -2,k_1+v-1,k_2 + s-1)$, which arise by relabeling $a \rightarrow a + \omega$ and $a \rightarrow a + A$ respectively in Eq.~\eqref{effAc}.

\subsubsection{Partial rotations}

Next, we consider the $4$-fold rotational symmetry operator $\hat{C}_4$, and we define $\hat{C}_{4,l} \equiv e^{i \frac{2\pi l}{4} \hat{N}} \hat{C}_4$ for integer $l$, with $\hat{N}$ the total $U(1)$ number operator.
We pick a 4-fold symmetric subregion $D$ whose length $L \gg \xi$, with $\xi$ the correlation length. Our main result, derived in general in Sec. \ref{sec:generalboson}, is that the rotation restricted to $D$, $\hat{C}_{4,l}|_D$, 
satisfies
\begin{align}
\label{partialRotation}
\mathcal{T}^b\left(\frac{2\pi}{4}; \frac{2\pi l}{4}\right) :=    \langle \Psi | \hat{C}_{4,l}|_D | \Psi \rangle  \approx e^{- \gamma |\partial D|} e^{i \frac{2\pi}{4} K_{l}} ,
\end{align} 
where $|\Psi\rangle$ is the ground state. Eq. \eqref{partialRotation} is expected to hold up to non-universal corrections exponentially small in $L/\xi$. 
Our theory predicts invariants $\Theta_l$ by taking appropriate modular reductions, assuming $D$ encloses an integer number of magnetic flux quanta:
\begin{equation}\label{eq:Theta-_vs_k3}
    \Theta_{l} := \begin{cases} K_l \mod 2, &\quad \text{if }  s + v l = 1 \mod 2 \\
    K_l \mod 4, &\quad \text{if } s + v l = 0 \mod 2\end{cases}
\end{equation}
These modular reductions can be understood in several ways. One is terms of certain redundancies in the G-crossed BTC description \cite{manjunath2020FQH,barkeshli2019} as explained in App.~\ref{app:relabel} and summarized by Eq.~\eqref{eq:bSET-relabel-U(1)_2}. The other is from a real space construction, which we explain in Sec.~\ref{sec:RealSpace}. We further find 
\begin{align}\label{eq:Kl}
K_l = -\frac{3}{4} +K_l^{\text{frac}} + K_l^{\text{SPT}} + A_l,
\end{align}
with $K_l^{\text{frac}} =    l^2 \frac{v^2}{4}+ l \frac{vs}{2} + \frac{s^2}{4} \mod 4$, $K_l^{\text{SPT}} = l^2 k_1  + l k_2 +  k_3 \mod 4$, and
$A_l = \frac{1}{4} \delta( [lv + s+ 1]_2) $,
where $\delta([x]_2)=1$ when $x=0$ mod 2, otherwise 0. 
Note that $K_l^{\text{SPT}}$ is an integer. 
By computing the above expectation value for generic $l$, together with $\sigma_H$, 
one can completely determine the crystalline invariants $v, s, k_1, k_2, k_3$. 

\subsubsection{Parton construction: projected wave functions and effective field theory}

We can obtain model ground state wave functions of these crystalline SET phases by utilizing the parton construction. We write the boson operator as $b = f_1 f_2$, which are fermionic partons. This introduces a $U(1)$ gauge symmetry, $f_1 \rightarrow e^{i\theta }f_1$, $f_2 \rightarrow e^{-i\theta} f_2$, with associated $U(1)$ gauge field $\alpha$. We assume a mean-field state where each fermionic parton 
forms a free fermion state with Chern number $C_1 = C_2 = 1$. The bosonic wave function is obtained by projecting the partons to the same location:
\begin{align}
\label{partonwfn}
    \Psi_b(\{\vec{r}_i\}) = \psi_{1}(\{\vec{r}_i\})\psi_{2}(\{\vec{r}_i\})
\end{align}

The many-body state of each parton is further labeled by crystalline invariants. With $G = U(1)^f \times C_4$ symmetry, the additional crystalline invariants form a $\mathbb{Z}_8 \times \mathbb{Z}_2$ classification and can be characterized by two invariants $(\mathscr{S}, \ell_s)$ \cite{manjunath2023classif}. These contribute terms $\frac{\mathscr{S}}{2\pi} A d \omega + \frac{\ell_s}{4\pi} \omega d \omega$ to the response theory. Below we will show that the invariants $v,s, k_{1}, k_{2}, k_3$ of the $1/2$ Laughlin state can be obtained from the crystalline invariants of the two parton states: $(\mathscr{S}_1, \ell_{s,1},\mathscr{S}_2, \ell_{s,2})$. We find:
\begin{align}
\label{partonPred}
v& = 1,\;\;\;\;\; s = \mathscr{S}_1 + \mathscr{S}_2,\;\; k_1 = k_2 = 0
    \nonumber \\
    k_3 &= \ell_{s,1} + \ell_{s,2} - \mathscr{S}_1^2 - \mathscr{S}_2^2.
\end{align}
A more general choice of invariants (e.g. $v = 0$ or any choice of $k_1, k_2$) requires a more sophisticated parton construction. To derive Eq.~\eqref{partonPred}, first note that each parton state is described by an effective field theory, $\mathcal{L}_{f} =  \frac{1}{4\pi} K_{IJ} a^I d a^J - \frac{v_f^I}{2\pi} A d a^I - \frac{s_f^I}{2\pi} \omega d a^I$, where repeated indices are implicitly summed. To describe an invertible fermionic state with Chern number $C = 1$ and $(\mathscr{S}, \ell_s)$, we take $K = 1 \oplus \sigma_x$, $v_f = (1, 0, 0)$, $s_f = (\mathscr{S}, \ell_s - \mathscr{S}^2, 1)$, where $\sigma_x$ is a Pauli matrix. The effective field theory of the bosonic state then is obtained by coupling the two parton theories (labeled with indices $i = 1,2$) to the $U(1)$ gauge field $\alpha$:
\begin{align}
    \mathcal{L} =& \frac{1}{4\pi} K_{IJ} a^{I}_{(i)} d  a^{J}_{(i)} - \frac{s_{f,i}^I}{2\pi} \omega d  a^I_{(i)} - \frac{v_{f,1}^I}{2\pi} A d  a_{(1)}^{I}
    \nonumber \\
    &+ \frac{1}{2\pi} \alpha d  (v^I_{f,1} a_{(1)}^I - v^I_{f,2} a_{(2)}^I) .
\end{align}
Integrating out $\alpha$ enforces $a_{(1)}^1 = a_{(2)}^1 = a$ up to a gauge transformation. Integrating out the remaining gauge fields then gives Eq.~\eqref{effAc} with the couplings of Eq.~\eqref{partonPred}. Performing the construction with $C=-1$ parton states instead gives projected wave functions for $\U_{-2}$ topological order, see App.~\ref{sec:u1-2} for partial rotations in $U(1)_{-2}$.

\subsection{Including translational symmetry: $G = U(1) \times_\phi [\mathbb{Z}^2 \rtimes \mathbb{Z}_4]$}

\begin{figure}[t]
    \centering
    \includegraphics[width=7.5cm]{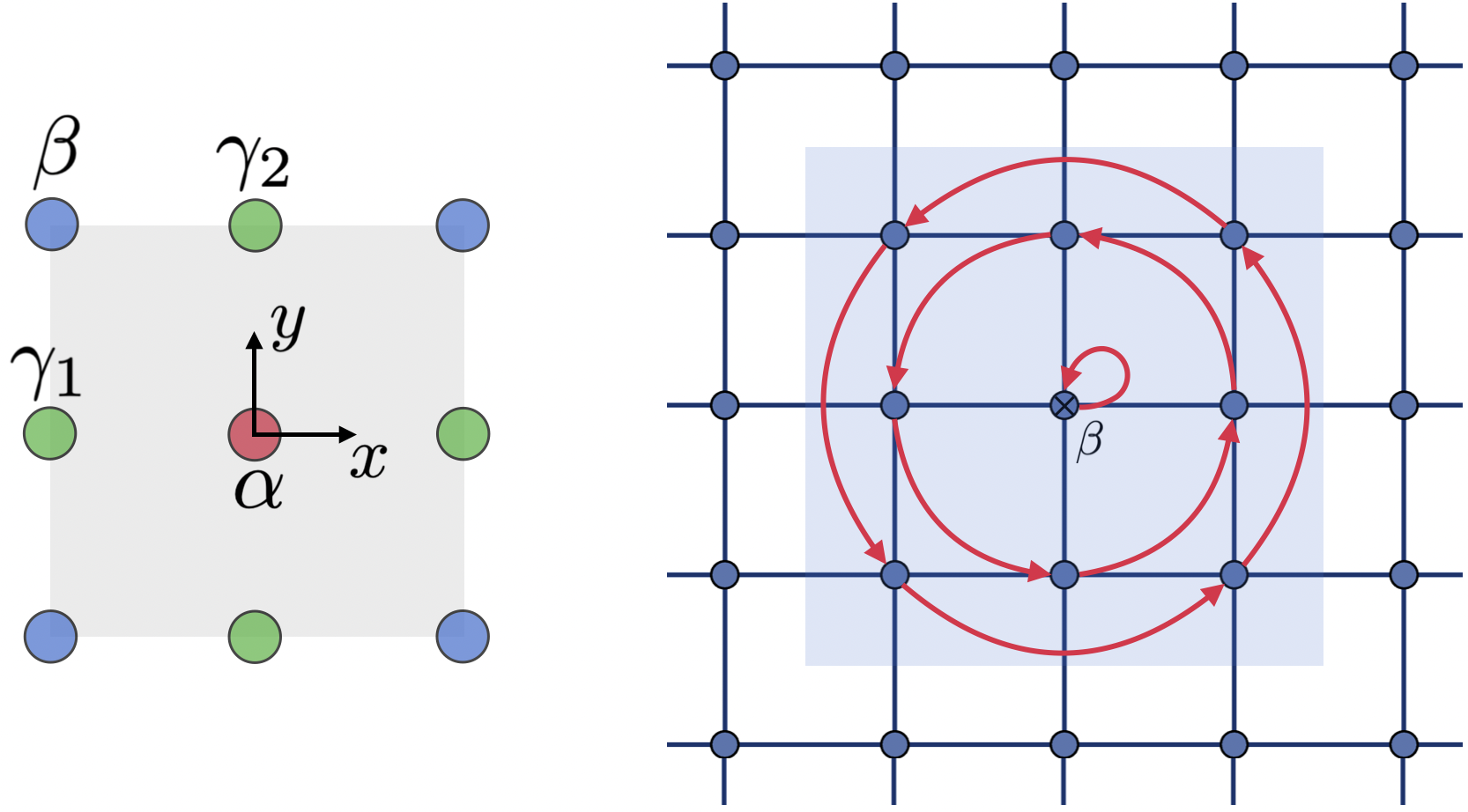}
    \caption{\textbf{Left.} Maximal Wyckoff positions for $C_4$ symmetry (colored circles).
    The high symmetry points $\gamma_1,\gamma_2$ both belong to the $\gamma$ maximal Wyckoff position, but are inequivalent under lattice translations. In our convention, the sites and plaquette centers are at $\beta$ and $\alpha$ respectively. \textbf{Right.} Partial rotation around $\OO=\beta$. The blue region is the partial rotation disk $D$, the red arrows represent the partial rotation operator $\hat{C}_{4,\beta}|_D$.}
    \label{fig:unit_cells}
\end{figure}

In the presence of translation symmetry and magnetic flux $\phi$ per unit cell, the symmetry becomes $G = U(1) \times_\phi [\mathbb{Z}^2 \rtimes \mathbb{Z}_4]$. We take  $\phi = 2\pi p /q$, with $p,q$ coprime. One can then consider rotations about distinct high symmetry points $\text{o}$ in the unit cell. The possible choices of $\text{o}$ are shown in Fig.~\ref{fig:unit_cells}. On a square lattice we have two high symmetry points with $C_4$ symmetry, which are the vertex and plaquette centers denoted $\text{o} = \alpha, \beta$ respectively. We also have the bond centers denoted $\text{o} = \gamma$ with $C_2$ symmetry. The rotational symmetry operator depends on $\text{o}$, so we have $\hat{C}_{4,\text{o}}$ and $\hat{C}_{4,\text{o},l}$ for $\text{o} = \alpha,\beta$, and $\pi$ rotations $\hat{C}_{2,\text{o}}$, with $\text{o} = \alpha,\beta,\gamma$, and $\hat{C}_{2,\text{o},l} := e^{i \pi \hat{N} l} \hat{C}_{2,\text{o}}$. The background gauge field $\omega$ also becomes origin dependent, $\omega_{\text{o}}$. $\omega_\alpha$, $\omega_\beta$ are $\mathbb{Z}_4$ gauge fields, $\omega_\gamma$ is a $\mathbb{Z}_2$ gauge field. 

For $\text{o} = \alpha, \beta$, we now have $\mathcal{T}^b_{\text{o}}(\frac{2\pi}{4}, \frac{2\pi l}{4})$ and $K_{\text{o},l}$, and $\Theta_{\text{o},l}$ defined analogously to Eqs.~\eqref{partialRotation},~\eqref{eq:Theta-_vs_k3}. 
For $\text{o} = \gamma$ we define $\mathcal{T}_{\gamma}^b(\pi, \pi l) := \langle (\hat{C}_{2,\gamma,l})_D \rangle = e^{-\gamma |\partial D|} e^{i \pi K_{\gamma,l}}$. 

Including translation symmetry has two main effects. First, the invariants that depend on rotational symmetry acquire an extra subscript $\text{o}$. Second, one can define a filling fraction (charge per unit cell) $\nu$. 
Specifically, symmetry fractionalization can be fully specified by $([v], [s_\alpha], [s_\beta], [s_\gamma])$. There is significant physics in the dependence of $[s_{\text{o}}]$ on $\text{o}$. In particular, there is an ``anyon per unit cell" $[m]$ characterizing the fractionalization of the translation algebra. There is also a discrete torsion anyon $[t_\alpha]$, which specifies the fractionalization of linear momentum and gives rise to a fractional quantized charge polarization \cite{manjunath2021cgt,manjunath2020FQH}. As we show in Sec.~\ref{sec:RealSpace} and App.~\ref{app:torsion}, these can be written as
\begin{align}
\label{mtsymfrac}
    [m] = [s_\alpha] \times [s_\beta] \times [s_\gamma]^2, \;\;\; [t_{\text{o}}] = [s_{\text{o}}] \times [s_\gamma], \; \OO = \alpha, \beta.
\end{align}

The filling fraction (charge per unit cell) $\nu$ obeys a Lieb-Schultz-Mattis type constraint \cite{Lu2020,manjunath2020FQH}
\begin{align}
    e^{2\pi i \nu} = e^{i 2\pi (vm/2 + \phi \sigma_H)}.
\end{align}
Fractional filling at zero flux $\phi$ requires $[v] = [m] = 1$.

A novel consequence of the above formulas is that when $\nu$ is fractional and the flux $\phi = 0$, then $[m]$ must be non-trivial, which in turn implies that at least one of $[s_{\text{o}}]$ must be non-trivial. Therefore a $1/2$ Laughlin FCI at zero field on a square lattice {\it necessarily must have fractionalization of the rotation symmetry about some high symmetry point $\text{o}$}. This is in addition to the well-known fractionalization of the translation algebra dictated by $[m]$. 

The SPT indices also become origin dependent, $k_{2,\text{o}}$ and $k_{3,\text{o}}$, for $\text{o} = \alpha, \beta, \gamma$. For $\text{o} = \alpha, \beta$, these are mod 4 invariants, while for $\text{o} = \gamma$ these are mod 2 invariants. The dependence of $k_{2,\text{o}}$ and $k_{3,\text{o}}$ on $\text{o}$ also contains significant physics, including a notion of angular momentum filling and angular momentum polarization~\cite{zhang2023complete,manjunath2023classif}. 

Our theory predicts the partial $C_2$ rotation around the origin $\gamma$ to be
\begin{align}
\mathcal{T}^b\left(\frac{2\pi}{2}; \frac{2\pi l}{2}\right)  \approx e^{- \gamma |\partial D|} e^{i \frac{2\pi}{2} K_{l,\gamma}} ,
\end{align}
with 
\begin{align}\label{eq:Klgamma}
K_{l,\gamma} = - \frac{1}{4} +K_{l,\gamma}^{\text{frac}} + K_{l,\gamma}^{\text{SPT}} + A_{l,\gamma},
\end{align}
where $K_{l,\gamma}^{\text{frac}} =    l^2 \frac{v^2}{4}+ l \frac{vs}{2} + \frac{s^2}{4} \mod 2$, $K_{l,\gamma}^{\text{SPT}} = l^2 k_1  + l k_2 +  k_3 \mod 2$, and
$A_{l,\gamma} = \frac{1}{4} \delta( [lv + s]_2) $.
Finally we have 
\begin{equation}\label{eq:Theta-gamma_vs_k}
    \Theta_{\gamma,l} := \begin{cases} K_{\gamma,l} \mod 1, &\quad \text{if }  s + v l = 0 \mod 2 \\
    K_{\gamma,l} \mod 2, &\quad \text{if } s + v l = 1 \mod 2\end{cases}.
\end{equation}
This modular reduction is also derived using Eq.~\eqref{eq:bSET-relabel-U(1)_2}.
The invariants defined above can now be used in the effective field theory. If we only consider the rotation gauge field $\omega_{\text{o}}$, we recover Eq.~\eqref{effAc}, just with $s, k_2, k_3, \omega$ replaced by $s_{\text{o}}$, $k_{2,\text{o}}$, $k_{3,\text{o}}$, $\omega_{\text{o}}$. Note that a more complete effective field theory would also include background translation gauge fields \cite{manjunath2021cgt,manjunath2020FQH} and additional coupling constants would directly specify the dependence of the invariants on $\text{o}$. 

Now we wish to describe the theory above in terms of the parton construction. Invertible fermionic states with $G = U(1) \times_\phi [\mathbb{Z}^2 \rtimes \mathbb{Z}_4]$ symmetry have a $\mathbb{Z}^3 \times \mathbb{Z}_8 \times \mathbb{Z}_2 \times \mathbb{Z}_4 \times \mathbb{Z}_4$ classification \cite{zhang2023complete,manjunath2023classif}. As shown in Ref.~\cite{manjunath2023classif}, the torsion part can be completely characterized by $\{\mathscr{S}_{\text{o}}, \ell_{s,\text{o}}\}$ for $\text{o} = \alpha, \beta, \gamma$. 
The results of the preceding section directly imply:
\begin{align}
    v& = 1,\;\;\;\;\; s_{\text{o}} = \mathscr{S}_{\text{o},1} + \mathscr{S}_{\text{o},2},\;\; k_1  =  k_{2,\text{o}} = 0 \nonumber \\
    k_{3,\text{o}} &= \ell_{s,\text{o},1} + \ell_{s,\text{o},2} - \mathscr{S}_{\text{o},1}^2 - \mathscr{S}_{\text{o},2}^2.
\end{align}
Note that for the projection to survive, each parton state must have the same filling, $\nu_p =  \frac{\phi_{p,i}}{2\pi} + \kappa_i$, for $i = 1,2$, where $\kappa_i$ is an integer, $\phi_{p,i}$ is the flux seen by each parton, and we have assumed each parton state has Chern number 1. The projection can survive only if the filling $\nu$ of the boson $b$ is equivalent to the filling of each parton state: $\nu = \nu_p$. Furthermore, $b$, being a composite of two partons, sees a flux per unit cell $\phi = \phi_{p,1} + \phi_{p,2}$. 
Therefore, $e^{2\pi i \nu} = e^{2\pi i (m v / 2 + \phi\sigma_H)}$, where $m = 2\nu-\frac{\phi}{2\pi}= \kappa_1+\kappa_2$.

\section{Numerical Monte Carlo results}

\begin{figure*}[t]
    \centering
    \includegraphics[width=16.5cm]{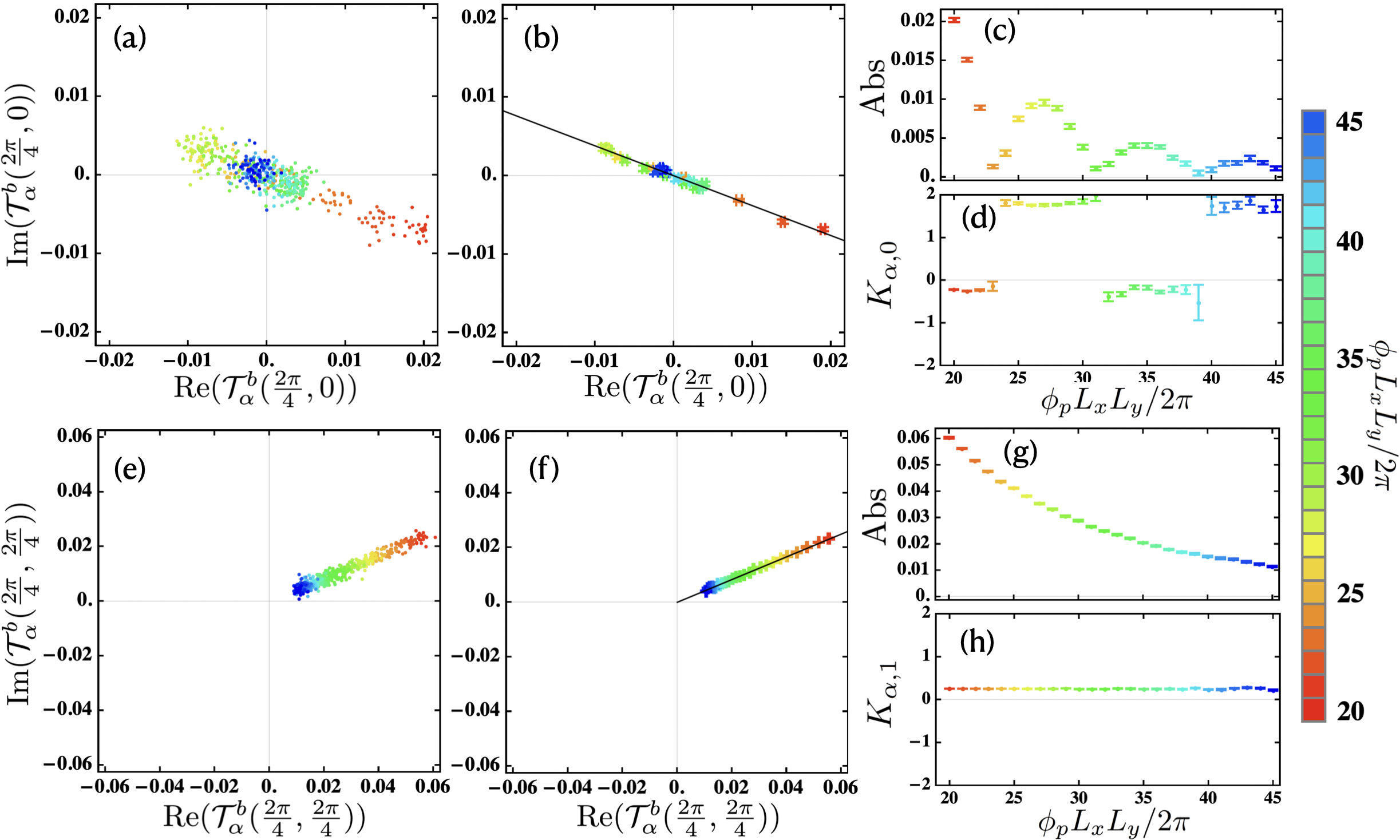}
    \caption{Raw data from QMC calculations. The system is defined on an $L_x \times L_y = 60\times 60$ torus with the partial rotation disk $D$ of size $30 \times 30$ and the rotation center $\OO=\alpha$ (plaquette center). We use the projected parton state with \{$C_1,\kappa_1,\mathscr{S}_{\alpha,1}, \ell_{s,\alpha,1}$\}=\{$C_2,\kappa_2,\mathscr{S}_{\alpha,2}, \ell_{s,\alpha,2}$\}=\{1,0,1/2,1/4\}.  \textbf{(a)} $l=0$. Each data point represents the expectation value $\mathcal{T}_\alpha^b(2\pi/M, 0)$ plotted in the complex plane and computed using $\approx 5 \times 10^6$ Metropolis-Hastings steps, referred to as a batch.  
    Points with the same color represent different batches with the same flux and filling. \textbf{(b)} Each data point is an average over 20 batches, with standard error shown as error bars. The slope $m$ of the fitted line gives $K_{\alpha,0}=\tan^{-1}(m)\times 4/(2\pi)$. \textbf{(c)(d)} $\text{Abs}(\mathcal{T}^b_\alpha(2\pi/M,0))$ and $K_{\alpha,0}$ for each flux. Note $K_{\alpha,0}$ can jump by $2$ as $\phi_p$ is varied. \textbf{(e,f,g,h)} are the same as \textbf{(a,b,c,d)} but with $l=1$. $K_{\alpha,1}$ does not show jumps. }
    \label{fig:raw_num}
\end{figure*}

\begin{table}[]
\renewcommand{\arraystretch}{1.5}
\centering
\begin{tabular}{c|c|c|c|c||c|c|c|c}
\multicolumn{9}{c}{Square Lattice $\U_{\pm 2}$ } \\
\hline \hline
&\multicolumn{4}{|c||}{$\ket{\psi_{C,\kappa}^2}$}   & \multicolumn{4}{c}{$\Theta_{\OO,l}$}
          \\ \hline 
$~\OO~$ & $~C~$ & $~\kappa~$ & $~\mathscr{S}_{\OO}~$ & $\ell_{s,\OO}$ & $~l=0~$ &  $~l=1~$ & $~l=2~$ & $~l=3~$ \\ 
\hline
\hline
$\beta$ & 1 & 0 & $1/2$ & 1/4 & -0.247 & 0.253 & -0.217 & -0.746 \\
\hline 
$\beta$ & -1 & 1 & $1/2$ & 15/4 & 0.236 & 0.750 & 0.227 & -0.251 \\
\hline 
$\alpha$ & 1 & 0 & $1/2$ & 1/4 &-0.232 & 0.251 & -0.227 & -0.750 \\
\hline 
$\alpha$ & -1 & 1 & $3/2$ & 7/4 & 0.208 & -0.251 & 0.228 & 0.748 \\
\hline 
$\alpha$ & 1 & -1 &$7/2$ & 1/4 & -0.220 & -0.749 & -0.233 & 0.253 \\
\hline 
$\alpha$ & -1 & 2 &$5/2$ & 7/4 & 0.245 & 0.747 & 0.218 & -0.251 \\
\hline 
$\alpha$ & 1 & -2 &$5/2$ & 9/4 & -0.233 & 0.252 & -0.223 & -0.750 \\
\hline 
$\alpha$ & -1 & 3 & $7/2$ & 15/4 & 0.213 & -0.249 & 0.241 & 0.748 \\
\hline 
$\alpha$ & 1 & -3 & $3/2$ & 9/4 & -0.226 & -0.748 & -0.227 & 0.250 \\
\hline 
$\alpha$ & -1 & 4 & $1/2$ & 15/4 & 0.231 & 0.750 & 0.211 & -0.251  \\
\hline 
$\gamma$ & 1 & 0 & $1/2$ & 1/4 & 0.000 & 0.000 & N/A & N/A \\
\hline
$\gamma$ & -1 & 1 & $1/2$ & 7/4 & 0.000 & 0.000 & N/A & N/A\\
\end{tabular}
\caption{QMC results of $\Theta_{\OO,l}$ for the squared projected parton ground states $\ket{\psi_{C,\kappa}^2}$, which denote the cases $\{C_1, \kappa_1\} = \{C_2, \kappa_2\}$.  For $\OO=\alpha,\beta$, $\Theta_{\OO,l}$ is defined modulo 2 when $l$ is even, and mod 4 when $l$ is odd. $\Theta_{\gamma,0}$ is defined mod 2, and $\Theta_{\gamma,1}$ is defined mod 1.
Values of $\mathscr{S}_{\OO}$ for the Hofstadter states are calculated using methods in \cite{zhang2022pol}. Values of $\ell_{s,\OO}$ for the Hofstadter states are calculated in \cite{zhang2023complete}. 
}\label{tab:theta1}
\end{table}

\begin{table*}[]
\renewcommand{\arraystretch}{1.5}
\centering
\begin{tabular}{c|c|c|c||c|c|c|c||c|c|c|c}
\multicolumn{12}{c}{Square Lattice $\U_{2}$ } \\
\hline \hline
\multicolumn{4}{c||}{$\ket{\psi_{C_1,\kappa_1}}$}&
\multicolumn{4}{c||}{$\ket{\psi_{C_2,\kappa_2}}$}& \multicolumn{4}{c}{$\Theta_{\alpha,l}$}
          \\ \hline 
 $C_1$ & $\kappa_1$ & $\mathscr{S}_{\alpha,1}$ & $\ell_{s,\alpha,1}$& $C_2$ & $\kappa_2$ & $\mathscr{S}_{\alpha,2}$ & $\ell_{s,\alpha,2}$ & $l=0$ &  $l=1$ & $l=2$ & $l=3$ \\ 
\hline
\hline
1 & 0 & 1/2 & 1/4 & 1 & -1 & 7/2 &1/4& $-0.751\mod 4$ & $-0.228\mod 2$ & $0.253\mod 4$ &$-0.235\mod 2$\\
\hline
1 & 0 & 1/2 &1/4& 1 & -2 & 5/2 &9/4& $-0.230\mod 2$ & $-0.745\mod 4$ & $-0.226\mod 2$ &$0.252\mod 4$\\
\hline
1 & 0 & 1/2 &1/4& 1 & -3 & 3/2 &9/4& $0.252\mod 4$ & $-0.223\mod 2$ & $-0.749\mod 4$ &$-0.230\mod 2$\\
\hline
1 & -1 & 7/2 &1/4& 1 & -2 & 5/2 &9/4& $0.255\mod 4$ & $-0.215\mod 2$ & $-0.749\mod 4$ &$-0.224\mod 2$\\
\hline
1 & -1 & 7/2 &1/4& 1 & -3 & 3/2 &9/4& $-0.229\mod 2$ & $0.251\mod 4$ & $-0.217\mod 2$ &$-0.747\mod 4$\\
\hline
1 & -2 & 5/2 &9/4& 1 & -3 & 3/2 &9/4& $-0.747\mod 4$ & $-0.229\mod 2$ & $0.251\mod 4$ &$-0.231\mod 2$\\

\hline 
\end{tabular}
\caption{QMC result of $\Theta_{\alpha,l}$ for the projected parton ground states $\ket{\psi_{C_1,\kappa_1}}\ket{\psi_{C_2,\kappa_2}}$. $\mathscr{S}_{\OO},\ell_{s,\OO}$ are calculated as stated in Table~\ref{tab:theta1}. The theory
in Eq.~\eqref{eq:Theta-_vs_k3}\eqref{eq:Kl} is aligned with the numerics and predict that $\Theta_{\alpha,l}$ to be the nearest quarter integer of this table.}\label{tab:theta2}
\end{table*}

We consider the parton wave function of Eq.~\eqref{partonwfn} and compute the partial rotation expectation value using numerical Monte Carlo calculations. For the parton wave functions, we use the ground state wave functions of the free fermion Harper-Hofstadter model on a square lattice with magnetic flux $\phi_p$ per unit cell. The states $\ket{\psi_{C,\kappa}}$ of this model are labeled by the Chern number $C$ and $\kappa\equiv \nu - C\phi_p/(2\pi)$, where $\nu$ is the filling, and these determine the other invariants $\{\mathscr{S}_{\OO}, \ell_{s,\OO}\}$ \cite{zhang2022fractional,zhang2022pol,zhang2023complete}.

In Fig.~\ref{fig:raw_num} we show numerical results for $l=\{0,1\}$, $\OO=\alpha$. Both parton states have invariants $\{C,\mathscr{S}_{\alpha},\ell_{\alpha}\}=\{1,1/2,1/4\}$. $\frac{2\pi}{4}K_{\alpha,l}$ is then extracted from the partial rotation expectation value through Eq.~\eqref{partialRotation}. The complex phase of $\bra{\Psi}(\hat{C}_{4,l})_D\ket{\Psi}$ can jump by $\pi$ as $\phi_p$ is varied. For example, in Fig.~\ref{fig:raw_num} (b), $K_{\alpha,0}$ could be either $\approx-0.25 \mod 4$ or   $\approx1.75 \mod 4$ depending on the flux. Therefore, to get an invariant, we need to define $\Theta_{\alpha,0}=K_{\alpha,0} \mod 2$ in this case.  
In Fig.~\ref{fig:raw_num} (e), $K_{\alpha,1} \approx 0.25 \mod 4$ for all flux $\phi_p$. Consequently, $\Theta_{\alpha,1}=K_{\alpha,1} \mod 4$ is an invariant. These are consistent with the modular reductions predicted by our theory in Sec.~\ref{sec:laughlin}. 

The calculation has two sources of error. The first is the Monte Carlo error, shown as the error bars in Fig.~\ref{fig:raw_num}(b,f), and which can be reduced by taking more Monte Carlo steps. The other source of error is finite size effects, which causes $\arg( \mathcal{T}_{\OO}^b )$ to deviate from the theoretical prediction. 

We perform the above calculations using several different parton states that realize $\U_{\pm 2}$ topological order. For projected wave functions constructed with two identical parton states, we summarize the value of $\Theta_{\OO,l}$ for all $\OO=\alpha,\beta,\gamma$ and $l=0,1,2,3$ in Table ~\ref{tab:theta1}. For $C=1$ entries, the theoretical predictions are summarized in the preceding section.
For the $C=-1$ entries, the theoretical predictions are given in App.~\ref{sec:u1-2}. 
In all cases the theory is aligned with the numerics and predict $\Theta_{\OO,l}$ to be the nearest quarter integer. The Monte Carlo sampling error is of the order of $0.001$, therefore the main source of error is expected to be due to finite size effects.

For projected wave functions constructed with two different parton states, we summarize the value of $\Theta_{\OO,l}$ for $\OO=\alpha$ and $l=0,1,2,3$ in Table \ref{tab:theta2}. The numerical values demonstrate remarkable agreement with the theoretical predictions. Additional details of the calculations are summarized in App.~\ref{sec:num_detail}.

\section{Real space construction}\label{sec:RealSpace}

To gain additional insight into the formulas above, here we present a model for each symmetry fractionalization pattern by utilizing a real space construction \cite{Huang2017,Else2019,zhang2022real,manjunath2023classif}.

Consider starting with the ground state of a $U(1)_2$ topological order with $[v] = 1$, but with trivial crystalline symmetry fractionalization. We can then construct a distinct symmetric state by placing an Abelian anyon $[s_{\text{o}}]$ at each high symmetry point $\text{o} = \alpha,\beta,\gamma$ of the square lattice unit cell. The triple $([s_{\alpha}],[s_{\beta}],[s_{\gamma}])$ now determines a new SET state, relative to the original state we started with. For $\U_{2}$ on the square lattice, we get a $(\mathbb{Z}_2)^3$ classification of the possible triples $([s_{\alpha}],[s_{\beta}],[s_{\gamma}])$. In general one needs to mod out by equivalences that move anyons between high symmetry points (see App.~\ref{app:Realspace}), but these are trivial for $\U_{2}$.

We can interpret a $2\pi$ disclination centered at $\text{o}$ as inducing the anyon $[s_{\text{o}}]$. To see this, consider the system to be at the surface of a cube. Each of the 8 corners can be interpreted as a $\pi/2$ disclination centered at $\text{o} = \beta$, so we effectively have $2$ $2\pi$ disclinations. For $F$ faces, there will be $F+2$ vertices, so we have two extra $[s_\beta]$ anyons as compared to being on flat space, for which the number of faces are in one-to-one correspondence. In this sense, we can assign an anyon $[s_\beta]$ to a $2\pi$ disclination centered at $\beta$. Similar arguments apply for $\text{o} = \alpha$ and $\gamma$. 

 The results above agree with the same classification obtained in Refs. \cite{manjunath2021cgt,manjunath2020FQH} using a different approach based on crystalline gauge fields. In particular, for a given $C_4$ symmetric origin $\text{o}$, the classification is given up to equivalences in terms of an anyon $s_{\text{o}}$ (``spin vector"), a pair of anyons $\vec{t}_{\OO} = (t_{\OO,x},t_{\OO,y})$ (``torsion vector"), and an `anyon per unit cell', denoted $m$. The individual classifications of these anyons were given in Ref.~\cite{manjunath2021cgt}. In terms of the real space construction, $[m]$ is given in Eq. \ref{mtsymfrac}.
This follows because each unit cell has one $\alpha$ and $\beta$ site each, and two $\gamma$ sites.
Furthermore, the real-space construction shows that there can be a topological charge (anyonic) polarization. As discussed in App.~\ref{app:torsion}, this can be used to give a heuristic understanding of $\vec{t}_{\OO}$ in terms of differences of the form  $[s_{\OO + \hat{x}(\hat{y})} - s_{\OO}]$.

The real space construction helps us understand why a partial rotation centered at $\text{o}$ can distinguish $s_{\text{o}}$. Relative to the state with trivial crystal symmetry fractionalization, the partial $\hat{C}_{4,\text{o}}$ rotation in the state decorated with $\{[s_{\OO}]\}$ should pick up an extra phase of $e^{i \frac{2\pi}{4} h_{s_{\text{o}}}}$, where $h_{s_{\text{o}}} = s_{\text{o}}^2/4$ is the topological spin of $s_{\text{o}}$. 
To understand the partial $\hat{C}_{4,\text{o},l}$ rotation, note that $[s_{\text{o}}]$ can also be thought of as the anyon induced by four $C_4$ defects, which are lattice disclinations. Four $\hat{C}_{4,\text{o},l}$ symmetry defects differ from four $\hat{C}_{4,\text{o}}$ symmetry defects by a $2\pi l$ $U(1)$ flux, which induces $l$ copies of the vison $[v]$. Thus the result for $\hat{C}_{4,\text{o},l}$ can be obtained from the result for $\hat{C}_{4,\text{o}}$ by taking $[s_{\text{o}}] \rightarrow [s_{\text{o}} + l v]$, giving the phase $h_{s_{\text{o}} + l v} = s_{\text{o}}^2/4 + l v s_{\text{o}}/2 + l^2 v^2/4$. This explains the $K_{\text{frac}}$ contribution in Eq.~\eqref{eq:Kl}. The $K_{\text{SPT}}$ contribution can be understood by including a real space construction for bosonic SPTs \cite{Huang2017,Song2020}. The additional term involving $\delta( ..)$ requires a more sophisticated analysis. From the edge CFT computation presented in Sec.~\ref{sec:generalboson}, we can see that this is an additional contribution arising fundamentally because the entanglement spectrum is described by the CFT at high temperatures. 

The real space construction also shows why we need to take the modular reductions of Eq.~\eqref{eq:Theta-_vs_k3}. Suppose that we increase the size of the disk $D$ to include more anyons. Since $D$ has to be 4-fold symmetric, it can include an integer multiple of 4 copies of an anyon $[a]$. Now under the partial $C_{4,\text{o},l}$ rotation, we consider four $[a]$ anyons making a $2\pi/4$ rotation around $[s_{\text{o}}]$, which is equivalent to a single $[a]$ rotating by $2\pi$ around $[s_{\text{o}}]$, giving a phase $e^{2\pi i a s_{\text{o}}/2}$. Furthermore, if $[a]$ has a charge of $a v/2 \mod 1$, the $U(1)$ rotation by $2\pi l /4$ picks up an additional phase of $e^{2\pi i a v l/2}$. Therefore, we have an ambiguity $K_{l} \rightarrow K_l + 2 a s_{\text{o}} + 2 a v l$. To get an invariant for all possible $a$ then leads us to Eq.~\eqref{eq:Theta-_vs_k3}. A more formal explanation is given in Sec. \ref{sec:generalboson} and App.~\ref{app:relabel} based on relabeling symmetry defects in the $G$-crossed BTC description. 

\section{General bosonic topological orders}
\label{sec:generalboson}

Here we evaluate the ground state expectation value of the partial rotation by an angle $2\pi/M$ restricted to a disk subregion $D$. Our result applies to general SET states of bosons, including both Abelian and non-Abelian states and to the case where symmetries may permute anyon types. We assume the ground state in the disk has trivial topological charge, integer magnetic flux, and no lattice defects. 
The choice of high symmetry point $\text{o}$ is implicit. 

To perform the calculation, we utilize the correspondence between the entanglement spectrum and edge CFT \cite{Haldane2008entanglement,Qi2012entanglement}.
Let the length of the boundary be $L=|\partial D|$. The partial rotation is then evaluated as a translation by $\frac{L}{M}$ acting in the edge CFT.
This translation acts in the edge CFT by a combination of an internal $\Z_M$ symmetry and the Lorentz translation symmetry in the 1d space:
\begin{align}
\begin{split}
\mathcal{T}^b\left(\frac{2\pi}{M};\frac{2\pi l}{M}\right) &= \bra{\Psi}\hat C_{M,l}|_{D} \ket{\Psi} \\
   &= \frac{\mathrm{Tr}[e^{i \frac{2\pi l}{M} \hat{N}} e^{i\hat Q_M\frac{2\pi}{M}}e^{i\tilde{P}\frac{L}{M}}e^{-\xi H}]}{\mathrm{Tr}[e^{-\xi H}]},
    \label{eq:rotasCFT}
    \end{split}
\end{align}
where the trace is taken in the trivial sector of CFT. The internal $\Z_M$ symmetry is generated by the operator $\hat{Q}_M$. $H$ is the CFT Hamiltonian density $H=\frac{2\pi}{L}(L_0-c_-/24)$, and
$\tilde{P}$ is a (normalized) translation operator $\tilde P=2\pi L_0/L$. Note the CFT is effectively at a high temperature because $\xi/L \ll 1$ is effectively an inverse temperature. 
In App.~\ref{app:cft} we present a general formula for $\mathcal{T}^b$ in terms of G-crossed BTC data that characterizes the crystalline SET. In the special case where the $C_M$ symmetry does not permute anyons, 
the result simplifies to the following form:
\begin{align}\label{eq:partialrot_boson_nonpermute}
\begin{split}
   \mathcal{T}^b\left(\frac{2\pi}{M};\frac{2\pi l}{M}\right) \propto &{e^{-2\pi i (M + \frac{2}{M})\frac{c_-}{24}}}\mathcal{I}_{M,l}\sum_{a}d_a \theta_a^M e^{2\pi i Q_{a,l}}
   \\
   &\times \sum_{b} S_{ab} e^{\frac{2\pi i}{M} h_b} e^{-\frac{2\pi L}{M^2\xi}(h_b-\frac{c_-}{24})},
   \end{split}
   \end{align}
where $\propto$ means being proportional up to a real positive number.  We sum over all possible  anyons $a,b$.
$d_a$ is the quantum dimension of $a$, $\theta_a = e^{2\pi i h_a}$ is the topological twist of $a$, $h_a$ is the topological spin, and $S$ is the modular $S$-matrix. For the sum over the $b$, the trivial anyon has the leading order contribution, with the other terms exponentially suppressed. When the contribution of the trivial anyon vanishes, which happens in the $U(1)_2$ case, the leading term is determined by the non-trivial anyon $a$ with the smallest scaling dimension in the chiral edge CFT. We have also defined the quantities $\mathcal{I}_{M,l}$ 
and $Q_{a,l}$, which are invariants determined by the response of the $U(1)\times \Z_M$ symmetry.
Using some technical results from \cite{manjunath2020FQH}, in App.~\ref{app:cft} we show that
\begin{align}
    e^{2\pi i Q_{a,l}} := M_{a, s\times v^l},
\end{align}
\begin{align}
    \mathcal{I}_{M,l} = e^{\frac{2\pi i}{M}(k_3+h_s + l (k_2 + m_{v,s}) + l^2(k_1+h_v)) },
    \label{eq:IMtheta}
\end{align}
with $M_{a,b}$ the mutual braiding between anyons $a,b$. $v$ and $s$ are Abelian anyons; $v$ is the vison, which determines the fractional $U(1)$ charge of anyons, while $s$ is the generalization of the spin vector, which determines the fractional orbital angular momentum of anyons \cite{manjunath2020FQH}. $k_1, k_2, k_3$ are the integers parameterizing the $\mathcal{H}^3(G, U(1))$ freedom of the SET \cite{barkeshli2019}. In the case of $U(1)_2$ CS theory, the sum over $a,b$ in Eq.~\eqref{eq:partialrot_boson_nonpermute} accounts for the additional $A_l$ term involving $\delta(...)$ in Eq.~\eqref{eq:Kl}. In App.~\ref{app:kmatrix}, we present a simplification of Eq.~\eqref{eq:IMtheta} for general Abelian topological orders using the $K$-matrix of Abelian CS theory. 

Finally, $\mathcal{T}^b$ only gives an invariant modulo a $U(1)$ phase. There are 
certain equivalences on $\mathcal{I}_M$ obtained by relabelling symmetry defects with anyons. As we explain in App.~\ref{app:relabel}, such relabellings induce the transformation
\begin{equation}\label{eq:bSET-relabel-U(1)_2}
    \mathcal{I}_{M,l} \rightarrow \mathcal{I}_{M,l} \times M_{s\times v^l,a} \times \theta_a^M,
\end{equation}
for any anyon $a$. 
From this result we recover the modular reductions of Eqs.~\eqref{eq:Theta-_vs_k3} and~\eqref{eq:Theta-gamma_vs_k}, which give the invariants $\Theta_{\OO,l}$.

We note that the results above also apply to partial rotations in continuum fractional quantum Hall states with continuous $SO(2)$ rotational symmetry, under the assumption of large rotation angles, $2\pi/M \gg 2\pi \sqrt{\xi/L}$.
For small rotation angles, $2\pi/M \ll 2\pi \sqrt{\xi/L}$,
a different analysis is required, as discussed in App.~\ref{smallangles}. 

The above general theoretical calculation assumes the correspondence between the entanglement spectrum and the edge CFT. In general, the entanglement spectrum also receives non-universal contributions, and it is an important direction to study the extent to which such non-universal contributions can cause deviations from the theory presented above. 

\section{Acknowledgements}

We thank Brayden Ware for collaboration during the initial stages of this work. 
This work is supported by NSF DMR-2345644 and the Laboratory for Physical Sciences through the Condensed Matter Theory Center. Research at Perimeter Institute is supported in part by the Government of Canada through the Department of Innovation, Science and Economic Development and by the Province of Ontario through the Ministry of Colleges and Universities.

\bibliography{bibliography.bib}

\newpage 

\appendix

\section{Numerical Details}\label{sec:num_detail}

\subsection{Review: Partial Rotations in Hofstadter Model}

\begin{figure*}[t]
    \centering
    \includegraphics[width=15cm]{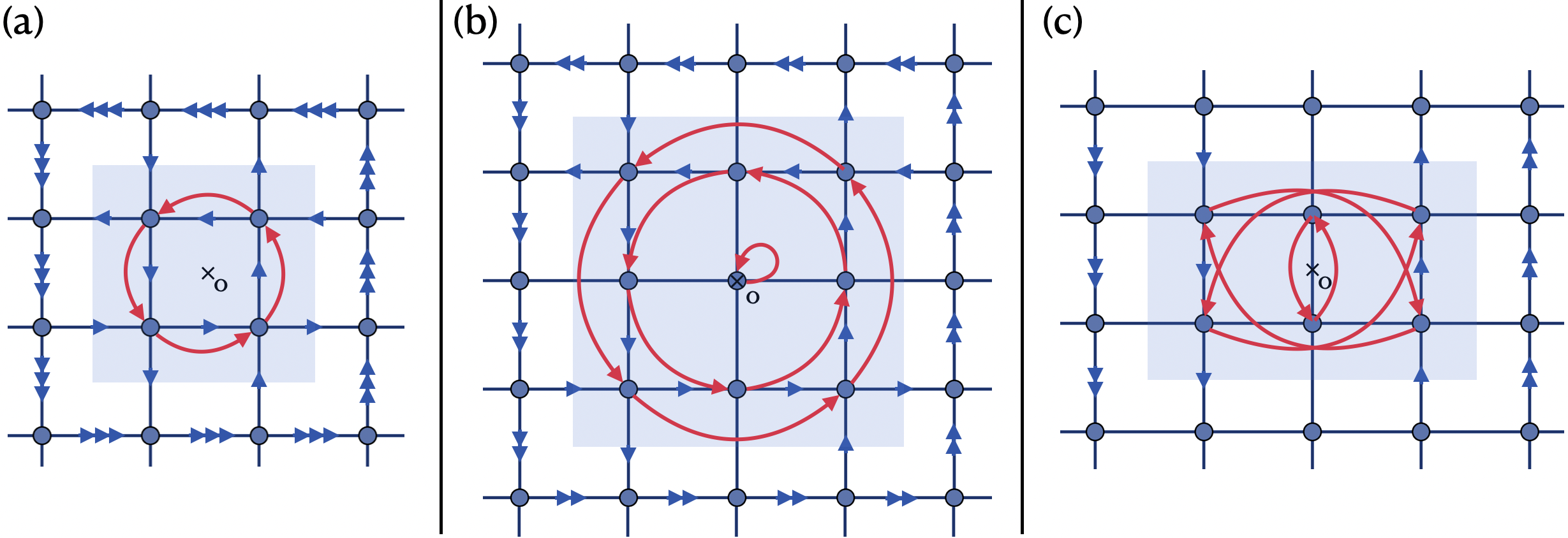}
    \caption{Square lattice Hofstadter model around the partial rotation disk (blue region) for different origins. The blue dots represent sites, the bonds represent hoppings, and the blue arrows represents the vector potential $A_{ij}$. We insert $\phi_p$ flux per plaquette. (a) $\OO=\alpha$ and each blue arrow represents a hopping phase of $A_{ij}=\phi_p/4$, (b) $\OO=\beta$ and each blue arrow represents a hopping phase of $A_{ij}=\phi_p/2$, (a) $\OO=\gamma$ and each arrow represents a hopping phase of $A_{ij}=\phi_p$. Under these symmetric gauges, the canonical partial rotation operators $\tilde{C}_{M_{\OO}}$ reduce to the bare partial rotation operator $\tilde{C}_{M_{\OO}}$ which does not involve any $\U$ phase. $\tilde{C}_{M_{\OO}}$ is just a permutation of sites shown as the red arrows.}
    \label{fig:partial_rotations}
\end{figure*}

In this section, we review the definition of the rotation operators and the methods in \cite{zhang2023complete} which are used to calculate the partial rotation invariant $\Theta_{\OO,l}$ and the associated field theory coefficient $\ell_{s,\OO}$ in the free fermion Hofstadter model. Note that $\ell_{s,\OO}$ is needed to compute the invariants for the parton mean-field states that are projected to give FCI wave functions.

As shown in \cite{zhang2022fractional,zhang2022pol,zhang2023complete}, the magnetic point group rotation operator is defined as $\hat{C}_{M_{\OO}}\equiv e^{\sum_j i\lambda_j \hat{N}_j} \hat{C}_{M_{\OO}}^{\text{bare}}$, where $\hat{C}_{M_{\OO}}^{\text{bare}}$ is a many body rotation operator which only moves points, and $\lambda_j$ is the $\U$ gauge transformation at site $j$ required to make the operator commute with the Hamiltonian. 
$M_{\OO}$ is the order of the point group symmetry at $\OO$. On the square lattice, $M_{\alpha}=M_{\beta}=4, M_{\gamma}=2$.
There are $M_{\OO}$ different choices of $\hat{C}_{M_{\OO}}$ that satisfy $(\hat{C}_{M_{\OO}})^{M_{\OO}}= + 1$. Among these, there is a canonical choice we denote $\hat{C}_{M_{\OO}}^+$; a disclination created using this operator does not carry any extra magnetic flux at the disclination core. The partial rotation operator $\hat{C}_{M_{\OO}}^+|_D$ is defined as $\hat{C}_{M_{\OO}}^+$ restricted to the disk $D$ and acting as an identity outside of $D$.

The Hofstadter model is defined by the Hamiltonian:

\begin{equation}
    H = -t\sum_{<ij>} e^{-i A_{ij}} c_i^{\dagger} c_j + h.c.
\end{equation}
with $i,j$ site indexes. The vector potential $A_{ij}$ threads $\phi_p$ flux per plaquette.
For each different origin, we can always find a choice of symmetric vector potential $A_{ij}$ such that the gauge transformation is unnecessary and $\hat{C}_{M_{\OO}}^+|_D$ reduces to the bare partial rotation $\hat{C}_{M_{\OO}}^{\text{bare}}|_D$. The choices of $A_{ij}$ we use are shown in Fig.~\ref{fig:partial_rotations}. 
The expectation value of $\hat{C}_{M_{\OO}}^{+}|_D$ for Hofstadter states $\ket{\Psi}$ is defined similar to Eq.~\eqref{partialRotation}:
\begin{align}\label{app:partialRotation}
\langle \Psi | \hat{C}_{M_{\OO}}^{+}|_D | \Psi \rangle  \approx e^{- \gamma |\partial D|} e^{i \frac{2\pi}{4} K_{l}}.
\end{align} 

The invariants $\Theta_{\OO}^+$ satisfy the following modular reduction:

\begin{align}
    \Theta_{\OO}^+ :=
        K^+_{\OO} \mod \frac{M_{\OO}}{2}.
\end{align}

\begin{figure*}[t]
    \centering
    \includegraphics[width=15cm]{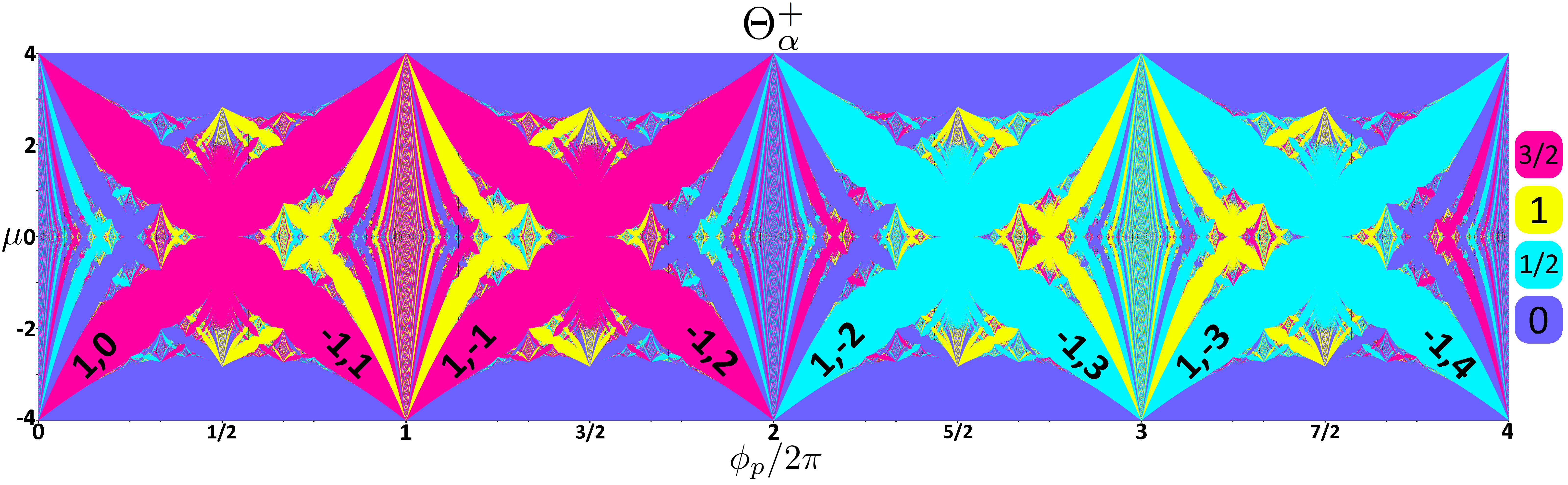}
    \caption{$\Theta_{\alpha}^+ \mod 2$ for Hofstadter model. $\{C,\kappa\}$ is labeled to the relevant lobes of the Monte Carlo calculation.}
    \label{fig:theta_alpha}
\end{figure*}

As an example $\Theta_{\alpha}^+$ is shown in Fig.~\ref{fig:theta_alpha} (incorporated from \cite{zhang2023complete}). We label $C$ and $\kappa$ in Fig.~\ref{fig:theta_alpha} to indicate the relevant bands which will be used in the Monte Carlo calculation.

On the square lattice $\ell_{s,\OO}$ is related to $\Theta_\OO^+$ through the following equation\footnote{Here, we suppressed the $+$ superscript of $\ell_{s,\OO}^+$ and corrected an error in Eq. 12 of \cite{zhang2023complete}. The correct equation is 
\begin{align*}
    \ell^{\pm}_{s,\OO} &=\begin{cases}
        \frac{11 \mp 1}{8}C+2\Theta^{\pm}_{\OO} \mod 4 \quad \OO=\alpha,\beta\\
        \frac{3\mp1}{8}C+2\Theta^{\pm}_{\OO} \mod 2\quad \OO=\gamma .
    \end{cases}
\end{align*}}:

\begin{align}\label{eq:ells_vs_Theta}
    \ell_{s,\OO} &=\begin{cases}
        \frac{5}{4}C+2\Theta^{+}_{\OO} \mod 4, \quad \OO=\alpha,\beta\\
        \frac{1}{4}C+2\Theta^{+}_{\OO} \mod 2, \quad \OO=\gamma.
    \end{cases}
\end{align}

We use Eq.~\eqref{eq:ells_vs_Theta} along with the values in Fig.~\ref{fig:theta_alpha} to generate the $\ell_{s,\OO}$ entries in Table~\ref{tab:theta1},\ref{tab:theta2}. 

\subsection{Constructing FCI wave functions from Hofstadter model ground states}\label{sec:constructTO}
The Hofstadter model offers us a plethora of ground states classified by the topological invariants $\{c_-,C,\kappa,\Theta_{\OO}^{\pm}\}$, making it a useful starting point to build states with topological order. For $C=\pm 1$ states, $\{C,\kappa\}$ uniquely defines the topological phase of matter associated with a Hofstadter ground state (for higher  $|C|$ this is generally not true) and they are used to label the ground state as $\ket{\psi_{C,\kappa}}$.

In the $\U_{\pm 2}$ parton construction, suppose there are $N$ hard core bosons arranged in the configuration
$\ket{r_1,r_2,\dots r_{N}}$, with $r_i$ denoting their positions. In the $U(1)_{\pm 2}$ theory we consider the boson operator
\begin{align}    b^\dagger_{r_i}=f_{1,r_i}^\dagger f_{2,r_i}^\dagger,
\end{align}
such that
\begin{align}
\ket{r_1,r_2,\dots r_{N}}=b_{r_1}^\dagger b_{r_2}^\dagger \dots b_{r_{N}}^\dagger \ket{0}.
\end{align}

The bosonic many-body ground state wave function for a given configuration is defined by multiplying the free fermion wavefunctions for the individual partons (the projection implies that the partons which form $b^{\dagger}_{r_i}$ are forced to be at the same location $r_i$). For example, consider the case where the two partons have equal $C$ and $\kappa$. The above procedure gives the ``squared state" $\ket{\Psi}=\ket{\psi_{C,\kappa}^2}$:

\begin{align}
    \bra{\psi_{C,\kappa}^2}\ket{r_1,r_2\dots r_{N}}&=\left(\frac{1}{\sqrt{N!}}\text{Det}(\mathcal{M})\right)^2\\
    \mathcal{M}_{ij}&=\chi_{i}(r_j),
\end{align}
where $\chi_i$ is a single particle eigenstate for one of the parton states. $\chi_i$ has energy $\epsilon_i$ below the chemical potential. 

We could also consider the case where the first parton state is labeled by $\ket{\psi_{C_1,\kappa_1}}$, and the second parton state by $\ket{\psi_{C_2,\kappa_2}}$. The wave function for $\ket{\Psi}=\ket{\psi_{C_1,\kappa_1}\psi_{C_2,\kappa_2}}$ is
\begin{align}
    \bra{\psi_{C_1,\kappa_1}\psi_{C_2,\kappa_2}}\ket{r_1,r_2\dots r_{N}}&=\frac{1}{N!}\text{Det}(\mathcal{M}_1)\text{Det}(\mathcal{M}_2)\\
    \mathcal{M}_{1,ij}=\chi_{1,i}(r_j)\qquad &
    \mathcal{M}_{2,ij}=\chi_{2,i}(r_j),
\end{align}
where $\chi_{1,i}$ and $\chi_{2,i}$ are single particle eigenstates for the two parton states.

We also define 
\begin{align}
\hat{C}_{M_\OO,l}|_D :=(e^{\sum_j i\frac{2\pi l}{M_\OO}\hat{N}_j}\hat{C}_{M_\OO}^{+})|_D.
\end{align}
For the squared states, we have the amplitude

\begin{align}
    \nonumber&\bra{\psi_{C,\kappa}^2}\hat{C}_{M_\OO,l}|_D\ket{r_1,r_2\dots r_{N}}\\
    \nonumber&=\bra{\psi_{C,\kappa}^2}\ket{R(r_1),R(r_2)\dots R(r_{N})} e^{i\frac{2\pi l}{4}n_D}\\
    &=e^{i\frac{2\pi l}{4}n_D}(\frac{1}{\sqrt{N!}}\text{Det}(\mathcal{M}'))^2.\\
    \nonumber&\qquad \mathcal{M}'_{ij}=\chi_{i}\left(R(r_{j})\right)
\end{align}

Here, $R(r_{j})=r_j$ if $r_j\not\in D$, and $R(r_{j})=r_k$ if $r_j\in D$ and $r_j$ rotates into $r_k$ under a $M_\OO$ fold counterclockwise rotation around $\OO$. $n_D$ is the number of bosons within region $D$ in the configuration $\ket{r_1,r_2\dots r_{N}}$.

Similarly, for the ground states $\ket{\Psi}=\ket{\psi_{C_1,\kappa_1}\psi_{C_2,\kappa_2}}$,
\begin{align}
    \nonumber&\bra{\psi_{C_1,\kappa_1}\psi_{C_2,\kappa_2}}\hat{C}_{M_\OO,l}|_D\ket{r_1,r_2\dots r_{N}}\\
    &=e^{i\frac{2\pi l}{4}n_D}\frac{1}{N!}\text{Det}(\mathcal{M}'_1)\text{Det}(\mathcal{M}'_2).\\
    \nonumber
    &~ \mathcal{M}'_{1,ij}=\chi_{1,i}\left(R(r_{j})\right)\quad \mathcal{M}'_{2,ij}=\chi_{2,i}\left(R(r_{j})\right)
\end{align}

\subsection{Variational Quantum Monte Carlo}
In this section we introduce the variational Quantum Monte Carlo method to calculate $\bra{\Psi}\hat{C}_{M_\OO,l}|_D\ket{\Psi}$ for the bosonic ground state $\ket\Psi$ constructed in the section above. The expectation value is expressed as:

\begin{align}\label{eq:expectationVal}
    \nonumber &\bra{\Psi}\hat{C}_{M_\OO,l}|_D\ket{\Psi}\\
    \nonumber
    =& \sum_{r_1,r_2,\dots r_{N}} \langle \Psi | \hat{C}_{M_\OO,l}|_D | r_1,r_2,\dots r_{N} \rangle \langle r_1,r_2,\dots r_{N} | \Psi \rangle\\
    =&\sum_{r_1,r_2,\dots r_{N}} \frac{\bra{\Psi}\hat{C}_{M_\OO,l}|_D\ket{r_1,r_2,\dots r_{N}}}{\bra{\Psi}\ket{r_1,r_2,\dots r_{N}}}P(r_1,r_2,\dots r_{N}),
\end{align}
where $P(r_1,r_2,\dots r_{N}):= |\bra{\Psi}\ket{r_1,r_2,\dots r_{N}}|^2$ is the probability of the configuration $\ket{r_1,r_2,\dots r_{N}}$.
This expectation value can be sampled and calculated through the Metropolis-Hastings algorithm \cite{hastings2010measuring}, which is summarized the the following 3 steps:

\begin{enumerate}
    \item Begin with a random initial configuration $\ket{r_1,r_2\dots r_{N}}$
    \item Propose a new configuration $\ket{\tilde{r}_1,\tilde{r}_2\dots \tilde{r}_{N}}$ by moving one particle to a random unfilled position. \textit{Update} the configuration to this new configuration if 
    \begin{align}
        \nonumber \left|\frac{\bra{\Psi}\ket{\tilde{r}_1,\tilde{r}_2\dots \tilde{r}_{N}}}{\bra{\Psi}\ket{r_1,r_2,\dots r_{N}}}\right|^2>q,
    \end{align}
    where $q$ is a random real number from 0 to 1; \textit{Reject} the new configuration and revert to $\ket{r_1,r_2\dots r_{N}}$ if the above condition is false. 
    \item Repeat Step 2. The expectation value in Eq.~\eqref{eq:expectationVal} is $\frac{\bra{\Psi}\hat{C}_{M_\OO,l}|_D\ket{r_1,r_2,\dots r_{N}}}{\bra{\Psi}\ket{r_1,r_2,\dots r_{N}}}$ averaged over the trajectory (with the first few iterations removed as they are not converged to the desired distribution).
\end{enumerate}

\subsubsection{Parameters and discussions}

We now define the parameters used in the calculations. 
Based on the periodicity of the free fermion invariants $\Theta_\OO^\pm$ in $\phi_p$ \cite{zhang2023complete}, we set $C$ and $\kappa$ to the values shown in Table~\ref{tab:theta2}. We first diagonalize the Hofstadter model and obtain the parton states $\ket{\psi_{C_1,\kappa_1}}$, $\ket{\psi_{C_2,\kappa_2}}$ on an $L_x\times L_y$ torus.

The parameters $\phi_p, N$ for each parton state satisfy the following relations with $\kappa$ and charge per unit cell $\nu$:
\begin{align}\label{eq:kappa}
    \nu &= C\frac{\phi_p}{2\pi} + \kappa\\
  \implies  N &= (C\frac{\phi_p}{2\pi} + \kappa )L_xL_y.
\end{align}

For $\OO=\alpha$, we use $L_x\times L_y=60\times 60$, for $\OO=\beta$, $L_x\times L_y=62\times 62$, and for $\OO=\gamma$, $L_x\times L_y=62\times 60$. We choose these parameters so that the linear size of $D$ can be exactly half of the system size which empirically works well. Though we have checked that having slightly larger or smaller $D$ or performing the calculation on a open disk does not change $\Theta_{\OO,l}$ in a significant way. Note that when calculating on an open disk, instead of filling $n$ particles, we place the chemical potential $\mu$ in the middle of the band gap and fill every energy level below $\mu$. Filling more or less edge states by moving $\mu$ within the band gap does not change $\Theta_{\OO,l}$ significantly in the numerical calculation.

In the CFT calculation discussed below, $\mathcal{T}^b_{\OO}$ is a sum of terms containing different powers of $e^{- L/\xi}$, where the exponents are fixed by the scaling dimensions of the anyons in the CFT. 
$L/\xi$ is the ratio between the size of the partial rotation disk boundary $L = |\partial D|$ and the correlation length. In the `sparse' limit where the particle number is very small, $\xi$ is proportional to the magnetic length $l_{B}$ which is the only other length scale in the system. This means that for a fixed $L$, $L/\xi\propto 1/l_{B}\propto \sqrt{\phi} \propto \sqrt{N}$. For some choices of $l$, higher order CFT contributions (involving larger powers of $e^{-L/\xi}$) identically vanish, and in these cases the phase of $\mathcal{T}^b_{\OO}$ remains a constant when changing $L/\xi$. However, for general choices of $l$ higher order contributions are nonvanishing, and in these cases we need to pick $L/\xi$ to be large enough to suppress the higher-order terms and measure a quantized phase of $\mathcal{T}^b_{\OO}$. 
There is a numerical trade-off here: Though a larger $L/\xi$ suppresses higher order contributions, it also reduces the overall amplitude $\text{Abs}(\mathcal{T}^b_{\OO})$, thereby reducing the accuracy of $\text{Arg}(\mathcal{T}^b_{\OO})$ from the Monte Carlo calculation.  This is because we have errors $\sigma_{\text{Re}}, \sigma_{\text{Im}}$ of $\mathcal{T}^b_{\OO}$ in the imaginary plane dictated by the amount of Monte Carlo steps and the amount of batches we take. The angular error $\sigma_{\text{Arg}}\propto \frac{1}{\text{Abs}(\mathcal{T}^b_{\OO})}\sigma_{\text{Re}}$.

In our numerics for the partial rotation with $\OO=\alpha,\beta$, 
we choose to fill $N$ particles where
$20 \le N \le 45$; the corresponding $\phi_p$ is calculated through Eq.~\eqref{eq:kappa}. Empirically, we find this choice suppresses the higher order contributions from the CFT to a reasonable amount, and allows us to measure a quantized $K_{\OO,l}$. For the $C_2$ partial rotation calculation with $\OO=\gamma$, the amplitude of $\mathcal{T}^b_{\OO}(\pi,\pi)$ nearly vanishes in the filling range $20 \le N \le 45$, therefore we choose a more sparse limit with the filling range
$3 \le N\le 15$, where there is a finite $\text{Abs}\left(\mathcal{T}^b_{\OO}(\pi,\pi)\right)$.

We run step 2 of the Metropolis-Hastings calculation 5,000,000 times in 20 separate batches of calculations. For $\OO=\alpha$, $l=0,1$, the raw data of $\mathcal{T}_{\OO,l}^b$ are shown in Fig.~\ref{fig:raw_num}.

\subsection{Calculation of topological entanglement entropy}

\begin{figure}[t]
    \centering
    \includegraphics[width=7.5cm]{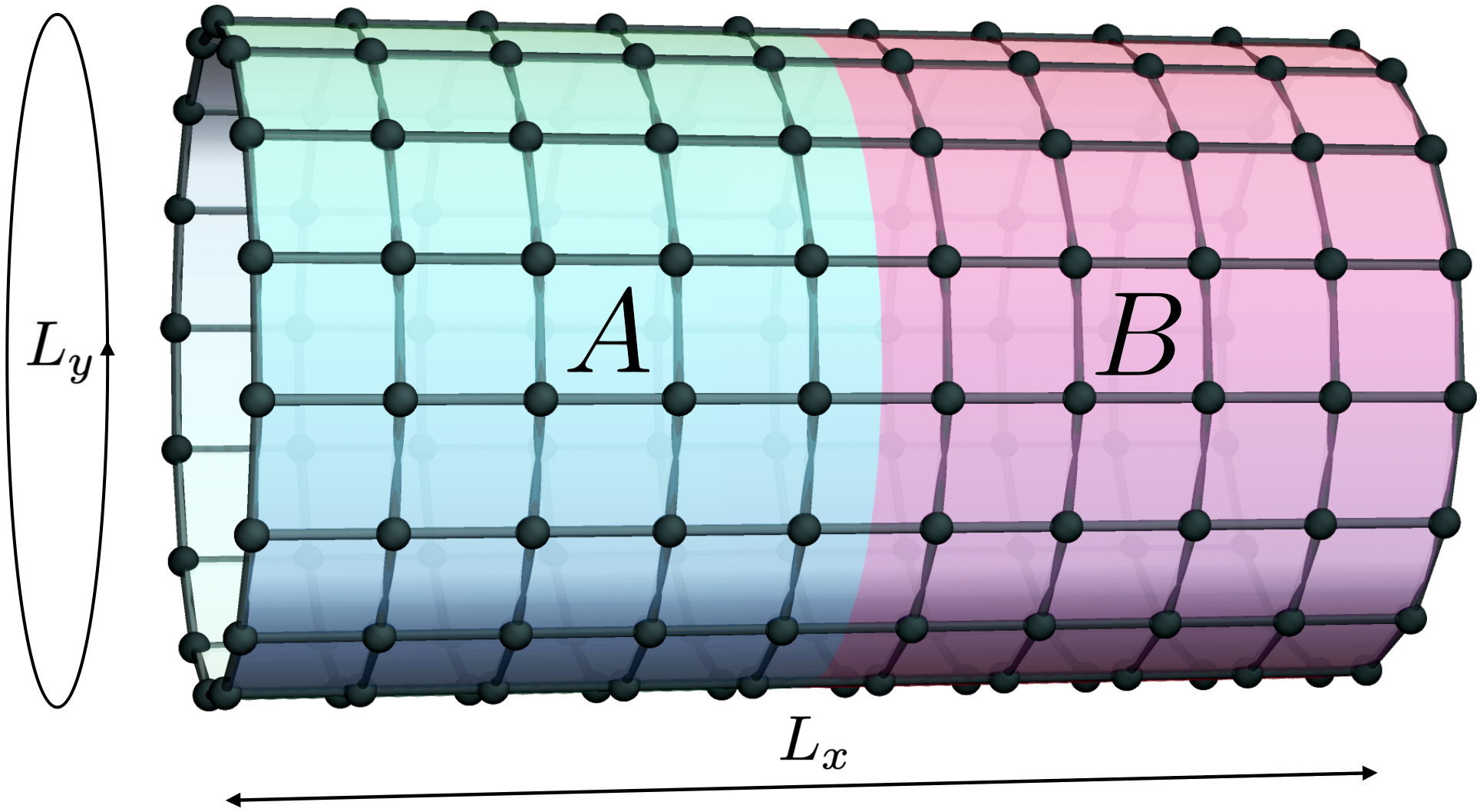}
    \caption{An $L_x\times L_y$ cylinder with periodic boundary condition in y direction bi-partitioned into region $A$ and $B$. This cylinder has $(L_x+1)\times L_y$ sites. }
    \label{fig:cylinder_bipartition}
\end{figure}

We calculate the topological entanglement entropy (TEE) to give evidence that our ground state $\ket{\Psi}$ from parton construction has topological order. Consider a bi-partition of a cylinder into regions $A$ and $B$ as shown in Fig.~\ref{fig:cylinder_bipartition}. Assuming $L_x$ is fixed, the Renyi entropy $S_2$ is predicted to follow\cite{kitaev2006topological,levin2006,jiang2012identifying} 
\begin{align}\label{eq:Renyi}
S_2=a L_y-S_{\text{topo}},
\end{align}
where $a$ is a constant. $S_{\text{topo}}$ is the TEE that encodes the total quantum dimension of the anyons. For a state without anyons, such as a integer quantum Hall state, $S_{\text{topo}}=0$, while for $\U_2$ topological order, $S_{\text{topo}} = \frac{1}{2} \text{ln}(2)\approx 0.347$. In this section, we numerically calculate $S_{\text{topo}}$ for a integer quantum Hall state and a squared parton state to show that the results are aligned with the prediction of Eq.~\eqref{eq:Renyi}.

We first prepare the Hofstadter state $\ket{\psi_{C,\kappa}}$ with $C=1$ and $\kappa = 0$ on a $L_x\times L_y$ cylinder (see Fig.~\ref{fig:cylinder_bipartition}), fixing $L_x=13$. We symmetrically partition the system into regions $A$ and $B$.

When defining the vector potential on the cylinder, we make sure that the total flux through the holes of the cylinder is trivial; that is, we demand that the holonomy computed at the two boundaries of the cylinder are integers. Since $\phi$ is a constant, this constrains the total flux $\phi_{\text{tot}}=\phi L_xL_y$ through the surface of the cylinder to be an integer multiple of $2\pi$, for any $L_y$. Thus we must have $\phi\in (2\pi/L_x) \Z$. We pick $\phi= 2\pi/L_x$ in our numerics and calculate $S_2$ for different $L_y$. 

We prepare two configurations $\ket{r_1,r_2,\dots r_n}$ and $\ket{r_1',r_2',\dots r_n'}$ on two different layers. The Renyi entropy can be calculated as expectation value of a partial layer swap operator \cite{ignacio2010infinite}, and is expressed as

\begin{align}
    &e^{-S_2}= \nonumber \\
    &\sum_{r_A, r_B,r_A',r_B'}\frac{\bra{\Psi}\ket{r_A,r_B'}\bra{\Psi}\ket{r_A',r_B}}{\bra{\Psi}\ket{r_A,r_B}\bra{\Psi}\ket{r_A',r_B'}}P(r_A,r_B)P(r_A',r_B')
\end{align}
where $r_A,r_B$ is shorthand for the set of electron positions in the $A,B$ subregions. We sample the probability $P(r_A,r_B)$ and $P(r_A,r_B)P(r_A',r_B')$ as before, but in each Metropolis-Hastings step we pick randomly whether to update $\ket{r_A,r_B}$ or $\ket{r_A',r_B'}$. Note that if the swapped configuration $\ket{r_A,r_B'}$ or $\ket{r_A',r_B}$ is not of the same filling as $\ket{\Psi}$, it is not in the projected Hilbert space and contributes nothing to $e^{-S_2}$.

\begin{figure}[t]
    \centering
    \includegraphics[width=7.5cm]{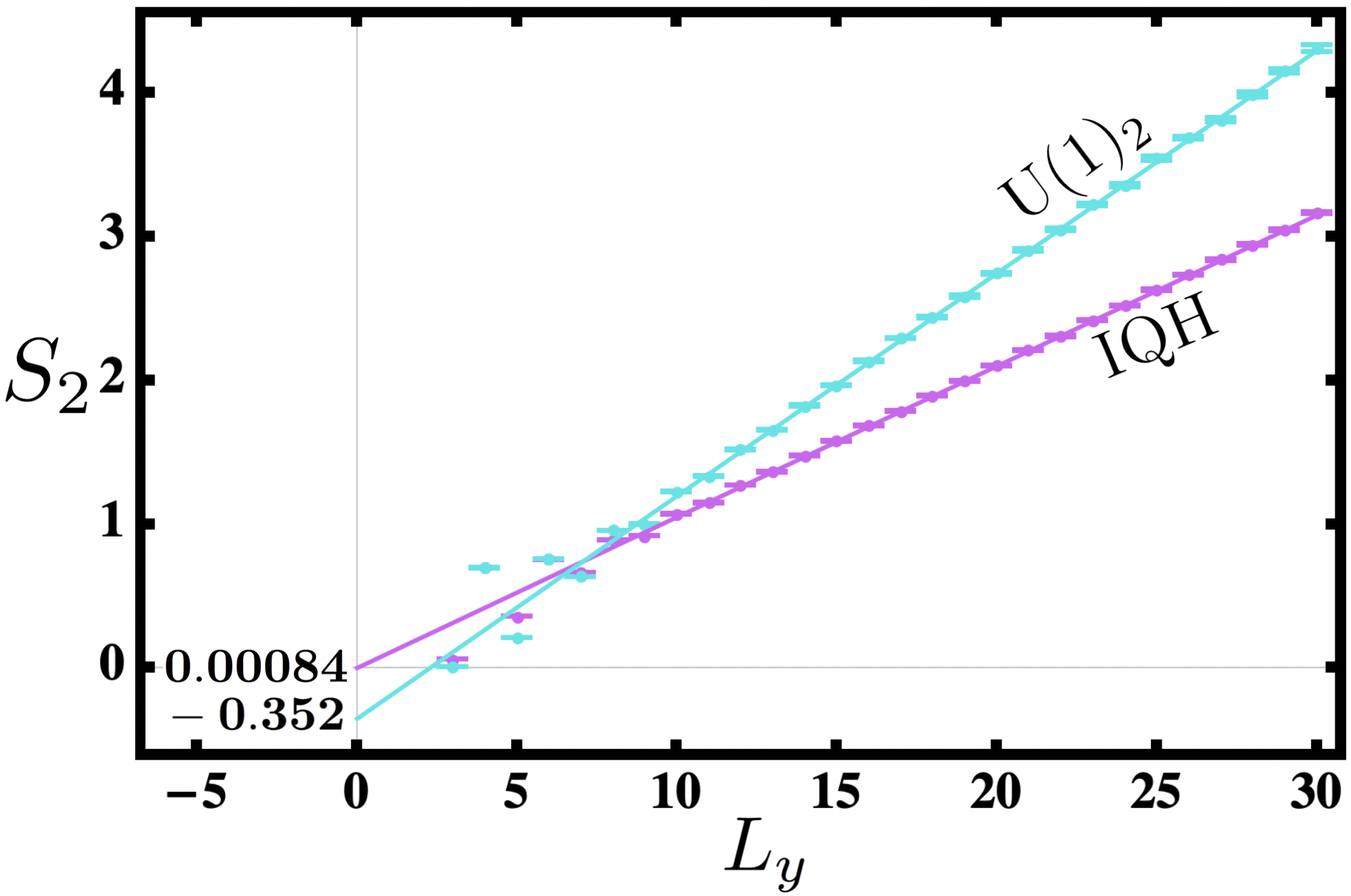}
    \caption{Renyi entropy $S_2$ for $L_x=14$ and different $L_y$ (error bars included). $S_2$ is fitted in the linear regime $14\le L_y\le 30$, and the TEE is calculated as the negative $L_y\rightarrow 0$ intercept}
    \label{fig:TEE}
\end{figure}

We perform the Monte Carlo calculation for the integer quantum Hall state $\Psi=\ket{\psi_{1,0}}$ and the squared state $\Psi=\ket{\psi_{1,0}^2}$ for $3\le L_y\le 30$. The resulting Renyi entropy is shown in Fig.~\ref{fig:TEE}. The TEE is extracted as the negative intercept, which demonstrates excellent agreement with the prediction Eq.~\eqref{eq:Renyi} ( within $1.6\%$ error). This gives further evidence that the ground states from parton construction are indeed 1/2 Laughlin states.

\section{The partial rotation in $U(1)_{-2}$ topological order}
\label{sec:u1-2}
In the main text, we mainly studied the partial rotation in the 1/2-Laughlin state described by the $U(1)_{2}$ topological order. 
Here we provide the formulae for partial rotations valid for the $U(1)_{-2}$ topological order, which we also compare to our numerical Monte Carlo calculations.
The partial rotation is given in the form
\begin{align}
\mathcal{T}^b\left(\frac{2\pi}{4}; \frac{2\pi l}{4}\right) :=    \langle \Psi | (\hat{C}_{4,l})_D | \Psi \rangle  \approx e^{- \gamma |\partial D|} e^{i \frac{2\pi}{4} K_{l}} ,
\end{align} 
with
\begin{align}\label{eq:Kl_-2}
K_l =  \frac{3}{4} -K_l^{\text{frac}} + K_l^{\text{SPT}} - A_l,
\end{align}
where $K_l^{\text{frac}} =    l^2 \frac{v^2}{4}+ l \frac{vs}{2} + \frac{s^2}{4} \mod 4$, $K_l^{\text{SPT}} = l^2 k_1  + l k_2 +  k_3 \mod 4$, and
$A_l = \frac{1}{4} \delta( [lv + s+ 1]_2) $,
where $\delta([x]_2)=1$ when $x=0$ mod 2, otherwise 0. 

The partial $C_2$ rotation along the bond center $\gamma$ is given by
\begin{align}
\mathcal{T}^b\left(\frac{2\pi}{2}; \frac{2\pi l}{2}\right)  \approx e^{- \gamma |\partial D|} e^{i \frac{2\pi}{2} K_{l,\gamma}} ,
\end{align}
with
\begin{align}\label{eq:Klgamma_-2}
K_{\gamma,l} = \frac{1}{4} - K_{l,\gamma}^{\text{frac}}  + K_{l,\gamma}^{\text{SPT}} - A_{l,\gamma},
\end{align}
with $K_{l,\gamma}^{\text{frac}} =    l^2 \frac{v^2}{4}+ l \frac{vs}{2} + \frac{s^2}{4} \mod 2$, $K_{l,\gamma}^{\text{SPT}} = l^2 k_1  + l k_2 +  k_3 \mod 2$, and
$A_{l,\gamma} = \frac{1}{4} \delta( [lv + s]_2) $.

When the $U(1)_{-2}$ topological order is formed by a pair partons forming Chern insulators with $C=-1$, its topological invariants are expressed in terms of those of the Chern insulators as
\begin{align}
    v& = 1,\;\;\;\;\; s_{\text{o}} = -\mathscr{S}_{\text{o},1} - \mathscr{S}_{\text{o},2},\;\; k_1  = k_{2,\text{o}} = 0
    \nonumber \nonumber \\
    k_{3,\text{o}} &= -\ell_{s,\text{o},1} - \ell_{s,\text{o},2} - \mathscr{S}_{\text{o},1}^2 - \mathscr{S}_{\text{o},2}^2.
\end{align}

\section{CFT calculations}
\label{app:cft}
\subsection{Derivation of partial rotation formula with anyon permutations}
\label{subsec:cftpermutation}
In this section, we perform the detailed CFT analysis outlined in Sec.~\ref{sec:generalboson}.
Following Ref.~\cite{Shiozaki2017point}, the quantity in Eq.~\eqref{eq:rotasCFT} is expressed in terms of CFT partition function on the edge. When $l=0$, the partial rotation is expressed as 
\begin{align}
    \begin{split}
    \mathcal{T}^b\left(\frac{2\pi}{M};0\right) &=
        \frac{\mathrm{Tr}[e^{i\hat{Q}_M\frac{2\pi}{M}}e^{i\tilde{P}\frac{L}{M}}e^{-\xi H}]}{\mathrm{Tr}[e^{-\xi H}]} \\
        &= e^{-\frac{2\pi i}{24M}c_-}\frac{\chi_1(\frac{i\xi}{L}-\frac{1}{M};0,1)}{\chi_1(\frac{i\xi}{L};0,0)}
        \end{split}
    \label{eq:rotcharacter_boson}
\end{align}
Here, $\chi_1(\tau;j,k)$ with $j,k\in\Z_M$ is the CFT character that corresponds to the partition function on a torus equipped with $\Z_M$ gauge field; $j,k$ denotes the $\Z_M$ twist along the spatial and temporal direction respectively. $\tau$ is the modular parameter and the subscript $1$ means the trivial sector of the Hilbert space. 

In the presence of the $\Z_M$ twisted boundary condition on the torus, the CFT character on the edge transforms under the modular $S,T$ transformations according to the $\Z_M$-crossed modularity of the $\Z_M$-crossed BTC~\cite{barkeshli2019}. To evaluate the CFT character, let us write the CFT character as 
\begin{align}
\begin{split}
\chi_1\left(\frac{i\xi}{L}-\frac{1}{M};0,1\right) &=\sum_b S_{1b}\chi_b\left(-\frac{1}{\frac{i\xi}{L}-\frac{1}{M}};1,0\right) \\
&= \sum_b(ST^M)_{1b}\chi_b\left(\frac{-iM\frac{\xi}{L}}{\frac{i\xi}{L}+\frac{1}{M}};1,0\right)
\label{eq:STM}
\end{split}
\end{align}
where $S,T$ are modular $S,T$ matrices of $\Z_M$-crossed BTC. They implicitly depend on the $\Z_M$ twist on a torus $(\mathbf{g},\mathbf{h})$ with $\mathbf{g},\mathbf{h}\in \Z_M$, and the $T$ matrix has the form of
\begin{align}
T^{(\mathbf{g},\mathbf{h})}_{a_{\mathbf{g}},b_{\mathbf{g}}} = e^{-\frac{2\pi i}{24}c_-}\cdot \theta_{a_{\mathrm{g}}}\cdot \eta_a(\mathbf{g},\mathbf{h})\cdot\delta_{a_{\mathbf{g}},b_{\mathbf{g}}} 
\end{align}
where the dependence on $(\mathbf{g},\mathbf{h})$ is made explicit.
The $\Z_M$-crossed BTC $\mathcal{C}_{\Z_M}$ has the structure of $\mathcal{C}_{\Z_M} = \bigoplus_{\mathbf{g}\in\Z_M} \mathcal{C}_{\mathbf{g}}$, and a simple object $a_{\mathbf{g}}\in\mathcal{C}_{\mathbf{g}}$ represents a vortex carrying $\mathbf{g}\in\Z_M$ in the bulk SET phase. The phase $\eta_b(\mathbf{g},\mathbf{h})$ describes the symmetry fractionalization of the vortex $b$~\cite{barkeshli2019}.
The vortex $b$ in Eq.~\eqref{eq:STM} carries the twist of $1\in\Z_M$ which is a generator of $\Z_M$, so $b\in\mathcal{C}_1$. 
Using the above $\Z_M$-crossed modularity, the character is further rewritten as
\begin{align}
   \begin{split}
&\chi_1\left(\frac{i\xi}{L}-\frac{1}{M};0,1\right) = \sum_{b\in\mathcal{C}_{1}}(ST^M)_{1b}\chi_b\left(\frac{-iM\frac{\xi}{L}}{\frac{i\xi}{L}+\frac{1}{M}};1,0\right) \\
&= e^{-\frac{2\pi iM}{24}c_-}\sum_{b\in\mathcal{C}_{1}} S_{1b} \theta_b^M \prod_{j=0}^{M-1} \eta_b\left(1,j\right)\cdot \chi_b\left(\frac{-iM\frac{\xi}{L}}{\frac{i\xi}{L}+\frac{1}{M}};1,0\right) \\
&= e^{-\frac{2\pi iM}{24}c_-}\sum_{b\in\mathcal{C}_{1}}\sum_{c\in\mathcal{C}_{0}} S_{1b} \theta_b^M \prod_{j=0}^{M-1} \eta_b\left(1,j\right) \\
&\quad \times S_{bc} \chi_c\left(\frac{iL}{M^2\xi}+\frac{1}{M};0,1\right)
\end{split} 
\end{align}
The above $ST^MS$ transformation makes the imaginary factor of the modular parameter $\tau=iL/(M^2\xi)+1/M$ large, assuming $\frac{L}{M^2\xi}\gg 1$. Recalling that $\chi_c=\mathrm{Tr}(e^{2\pi i \tau (L_0-\frac{c_-}{24})})$, the CFT character in the last expression can be approximated by
   \begin{align}
\begin{split}
    \chi_c\left(\frac{iL}{M^2\xi}+\frac{1}{M};0,1\right) &\approx e^{\frac{2\pi i}{M}(h_c-\frac{c_-}{24})} e^{-\frac{2\pi L}{M^2\xi}(h_c-\frac{c_-}{24})}
    \label{eq:character_approx}
    \end{split}
\end{align}
One can then express the partial rotation as
\begin{align}
   \mathcal{T}^b\left(\frac{2\pi}{M};0\right)  \propto & {e^{-2\pi i (M + \frac{2}{M})\frac{c_-}{24}}}\sum_{b\in\mathcal{C}_{1}}d_b \theta_b^M \prod_{j=0}^{M-1} \eta_b\left(1,j\right) 
   \nonumber \\
   &\times \sum_{c\in\mathcal{C}_0} S_{bc} e^{\frac{2\pi i}{M} h_c} e^{-\frac{2\pi L}{M^2\xi}(h_c-\frac{c_-}{24})}~.
\label{eq:partialrot_bosonic_intermediate}
   \end{align}
We note that this expression is valid even when the $C_M$ symmetry permutes anyons.

\subsection{Including partial $U(1)$ charge rotation}

In the presence of $U(1)$ symmetry with generic $l$, one can also evaluate the partial rotation associated with the partial $U(1)$ transformation within the disk,
\begin{align}
\begin{split}
\mathcal{T}^b\left(\frac{2\pi}{M};\frac{2\pi l}{M}\right) & :=
    \bra{\Psi} \hat C_{M,l}|_{D^2} \ket{\Psi} \\
    &= \frac{\mathrm{Tr}[e^{i\hat N\frac{2\pi l}{M}}e^{i\hat Q_M\frac{2\pi}{M}}e^{i\tilde{P}\frac{L}{M}}e^{-\xi H}]}{\mathrm{Tr}[e^{-\xi H}]}
    \label{eq:rotU1asCFT}
    \end{split}
\end{align}
The above quantity can also be computed by the same logic as above, and given by
\begin{widetext}
\begin{align}
   \mathcal{T}^b\left(\frac{2\pi}{M};\frac{2\pi l}{M}\right)  &\propto  {e^{-2\pi i (M + \frac{2}{M})\frac{c_-}{24}}}\sum_{b\in\mathcal{C}_{{2\pi l/M,1}}}d_b \theta_b^M 
   \prod_{j=0}^{M-1}\eta_{b}\left((\left[\frac{2\pi l}{M}\right]_{2\pi},1),(\left[\frac{2\pi jl}{M}\right]_{2\pi},j)\right) 
   \sum_{c\in\mathcal{C}_0} S_{bc} e^{\frac{2\pi i}{M} h_c} e^{-\frac{2\pi L}{M^2\xi}(h_c-\frac{c_-}{24})} 
   \label{eq:partialrot_boson_U(1)_permute}
   \end{align}
\end{widetext}
where $\mathcal{C}_{{\theta,1}}$ is the twisted sector with $(\theta,1)\in \U\times \Z_M$.

\subsection{Simplified expression for non-permuting $C_M$ symmetries}\label{app:IM,Lb}

The above formula \eqref{eq:partialrot_bosonic_intermediate} with $l=0$ is further simplified when the $C_M$ symmetry does not permute anyons.
Let us employ a general expression of these quantities valid for the case without permutation action~\cite{barkeshli2019},
\begin{align}
\begin{split}
    \theta_{b_{1}} &= \theta_{b_0}\theta_{0_{1}}\cdot (R^{b_0, 0_{1}}R^{0_{1}, b_0}), \\
    \eta_{b_1}\left(1,j\right) &= \eta_{b_0}\left(1,j\right)\eta_{0_1}\left(1,j\right), \\
    S_{a,b_{1}} &= \frac{1}{d_a}S_{a,b_0}(R^{a, 0_{-1}}R^{0_{-1}, a}).
    \label{eq:no_permutation_solutions}
\end{split}
\end{align}
Here $R^{ab}$ is the R-symbol of the G-crossed theory that specifies the algebraic braiding properties of anyons and defects \cite{barkeshli2019}. 
One can choose the gauge where $R^{b,0_{1}}=1$, $R^{c,0_1}=1$ for $b,c\in\mathcal{C}_0$. We then obtain
\begin{align}
\begin{split}
   \mathcal{T}^b\left(\frac{2\pi}{M};0\right) \propto &{e^{-2\pi i (M + \frac{2}{M})\frac{c_-}{24}}}\mathcal{I}_M \sum_{b\in\mathcal{C}_{0}}d_b \theta_b^M e^{2\pi i L_b} 
   \\
   &\sum_{c\in\mathcal{C}_0} S_{bc} e^{\frac{2\pi i}{M} h_c} e^{-\frac{2\pi L}{M^2\xi}(h_c-\frac{c_-}{24})}
   \label{eq:nonpermute}
   \end{split}
   \end{align}
   which shows Eq.~\eqref{eq:partialrot_boson_nonpermute}.
We defined the invariants
\begin{align}\label{eq:I_M,L_b-gen}
    \mathcal{I}_M &:= \theta_{0_{1}}^M\prod_{j=0}^{M-1}\eta_{0_1}\left(1,j\right), \nonumber \\
    e^{2\pi i L_b} &:= \prod_{j=0}^{M-1} \eta_b\left(1,j\right).
\end{align}
The $U(1)$ phase $\eta_a(\mathbf{g},\mathbf{h})$ describes the symmetry fractionalization of the topological charge $a$. $L_b$ is the fractional $\mathbb{Z}_M$ charge of $b$.
The gauge invariant quantity $\mathcal{I}_M$ is further computed by plugging the symmetry fractionalization data $\theta_{0_{\bf g}}, \eta_{0_{\bf g}}(\bf h, \bf k)$ into its expression, which was performed in~\cite{manjunath2020FQH}.
To do this, we can work in the specific gauge where $\theta_{0_{\bf g}}= 1$, and
\small
\begin{align}
    \begin{split}
        \eta_{0_{\bf g}}(\bf h, \bf k) &= \exp\left(2\pi i(h_s+k_3)\frac{[{\bf g}]_M}{M}\frac{[{\bf h}]_M+[{\bf k}]_M-[{\bf h}+{\bf k}]_M}{M}\right), 
    \end{split}
\end{align}
\normalsize
where we define $s\in\mathcal{C}_0$ as the relation $M_{b,s}= e^{2\pi i L_b}$ for all anyons $b\in\mathcal{C}_0$. The parameter $k_3\in\Z_M$ corresponds to the label of bosonic SPT phase with $C_M$ symmetry.
The quantity $\mathcal{I}_M$ is then computed as
\begin{align}
    \mathcal{I}_M = e^{\frac{2\pi i}{M}(h_s+k_3)}.
\end{align}

In the absence of the permutation action, the formula \eqref{eq:partialrot_boson_U(1)_permute} with generic $l$ can also be simplified by the same logic. It can be written as
\begin{align}
\begin{split}
   \mathcal{T}^b\left(\frac{2\pi}{M};\frac{2\pi l}{M}\right) \propto &{e^{-2\pi i (M + \frac{2}{M})\frac{c_-}{24}}}\mathcal{I}_{M,l}\sum_{a}d_a^2 \theta_a^M e^{2\pi i Q_{a,l}}
   \\
   &\times \sum_{b} S_{ab} e^{\frac{2\pi i}{M} h_b} e^{-\frac{2\pi L}{M^2\xi}(h_b-\frac{c_-}{24})},
   \label{eq:nonpermuteU1_app}
   \end{split}
   \end{align}
with 
\small
\begin{align}
    &\mathcal{I}_{M,l} := \theta_{0_{\frac{2\pi l}{M},1}}^M\prod_{j=0}^{M-1}\eta_{0_{\frac{2\pi l}{M},1}}\left((\left[\frac{2\pi l}{M}\right]_{2\pi},1),(\left[\frac{2\pi jl}{M}\right]_{2\pi},j)\right) , \nonumber \\
    &e^{2\pi i Q_{a,l}} := \prod_{j=0}^{M-1} \eta_b\left((\left[\frac{2\pi l}{M}\right]_{2\pi},1),(\left[\frac{2\pi jl}{M}\right]_{2\pi},j)\right),
\end{align}
\normalsize
which reduces to Eq.~\eqref{eq:I_M,L_b-gen} when $l=0$. By plugging the form of $\theta,\eta$ of defects in~\cite{manjunath2020FQH} into the above expression, we obtain
\begin{align}
    e^{2\pi i Q_{a,l}} := M_{a, s\times v^l},
\end{align}
\begin{align}
    \mathcal{I}_{M,l} = e^{\frac{2\pi i}{M}(k_3+h_s + l (k_2 + m_{v,s}) + l^2(k_1+h_v)) },
\end{align}
where $(k_1,k_2,k_3)\in \Z\times\Z_M\times\Z_M$ characterizes the bosonic SPT index, and $M_{v,s}=e^{2\pi i m_{v,s}}$.

\subsection{K-matrix theory}
\label{app:kmatrix}
For Abelian topological orders, we can use the more familiar language of K-matrix CS theory  coupled to a $\U\times\Z_M$ crystalline gauge field $(A,\omega)$:
\begin{align}
\begin{split}
    \mathcal{L} = & \frac{1}{4\pi} K_{IJ} a_I da_J - v_Ia_I \frac{dA}{2\pi}  - s_I a_I \frac{d\omega}{2\pi}\\
    &-  \frac{k_3}{2\pi} \omega d\omega - \frac{k_2}{2\pi} Ad\omega - \frac{k_1}{2\pi} AdA,
    \end{split}
\end{align}
where $k_1,k_2,k_3\in\Z$. $v_I, s_I$ are integer vectors that determine the fractionalization of $U(1)$ and $C_M$ symmetry respectively. In this case, when the leading contribution $a=0$ is not vanishing in Eq.~\eqref{eq:nonpermuteU1_app}, the partial rotation is given to leading order by
\begin{align}\label{eq:partialrot_Kmat}
\begin{split}
    & \mathcal{T}^b\left(\frac{2\pi}{M};\frac{2\pi l}{M}\right)  \\
    &\propto {e^{-2\pi i (M + \frac{2}{M})\frac{c_-}{24}}}e^{\frac{2\pi i}{M}(k_3+h_s + l (k_2 + m_{v,s}) + l^2(k_1+h_v)) } \\
    & \sum_{{\bf m}\in \Z^{|\bf K|}/\bf K\Z^{|\bf K|}} e^{M\pi i  {\bf m}^{\mathrm{T}}{\bf K}^{-1} {\bf m} }e^{2\pi i{\bf m}^{\mathrm{T}}{\bf K}^{-1} ({l \bf q+\bf s})}.
    \end{split}
\end{align}
In Tables~\ref{tab:U(1)_2-C4},~\ref{tab:U(1)_2-C2}, 
 we list the partial rotation invariants of bosonic $\U_2$ states with different symmetry fractionalization classes, which can be compared to the numerical results in Tables~\ref{tab:theta1},~\ref{tab:theta2}.

\begin{table}[t]
\renewcommand*{\arraystretch}{1.5}
\centering
\begin{tabular}{c|c|c| c}
\hline \hline
        $\U_2$ & $v$ & $s$   & $\mathcal{T}^b\left(\frac{2\pi}{4},\frac{2\pi l}{4}\right)$ \ ($l=0,1,2,3$) \\ \hline \hline
\ & $[0]_2$ & $[0]_2$ & $e^{-3\pi i/8},e^{-3\pi i/8},e^{-3\pi i/8},e^{-3\pi i/8}$ \\ \hline
\ & $[0]_2$ & $[1]_2$ & $e^{-\pi i/8},e^{-\pi i/8},e^{-\pi i/8},e^{-\pi i/8}$ \\ \hline
\ & $[1]_2$ & $[0]_2$ & $e^{-3\pi i/8},e^{-\pi i/8},e^{\pi i/8},e^{7\pi i/8}$ \\ \hline
\ & $[1]_2$ & $[1]_2$ & $e^{-\pi i/8},e^{\pi i/8},e^{7\pi i/8},e^{-3\pi i/8}$ \\ 
\hline \hline
\end{tabular}
\caption{$\mathcal{T}^b\left(\frac{2\pi}{4},\frac{2\pi l}{4}\right)$ for different symmetry fractionalization classes of $\U_2$, computed using Eq.~\eqref{eq:partialrot_Kmat}.}
\label{tab:U(1)_2-C4}
\end{table}

\begin{table}[t]
\renewcommand*{\arraystretch}{1.5}
\centering
\begin{tabular}{c|c|c| c}
\hline \hline
        $\U_2$ & $v$ & $s$   & $\mathcal{T}^b\left(\frac{2\pi}{2},\frac{2\pi l}{2}\right)$ \ ($l=0,1$) \\ \hline \hline
\ & $[0]_2$ & $[0]_2$ & $1,1$ \\ \hline
\ & $[0]_2$ & $[1]_2$ & $1,1$ \\ \hline
\ & $[1]_2$ & $[0]_2$ & $1, 1$ \\ \hline
\ & $[1]_2$ & $[1]_2$ & $1,-1$ \\ 
\hline \hline
\end{tabular}
\caption{$\mathcal{T}^b\left(\frac{2\pi}{2},\frac{2\pi l}{2}\right)$ for different symmetry fractionalization classes of $\U_2$, computed using Eq.~\eqref{eq:partialrot_Kmat}.}
\label{tab:U(1)_2-C2}
\end{table}

\subsection{Comparison between small and large rotation angle}
\label{smallangles}

In the main text, we mainly studied the partial rotation $\mathcal{T}^b(2\pi/M)$ where the rotation is associated with the point group symmetry of the lattice system with $M=1,2,3,4,6$. Meanwhile, if we instead consider a continuous system for topological order such as the Laughlin wave function, the system has continuous rotation symmetry and one can study the partial rotation $\mathcal{T}^b(\theta)$ with generic rotation angle $\theta\in\mathbb{R}/2\pi \Z$. 
In this appendix, we study the behavior of the partial rotation $\mathcal{T}^b(2\pi/M,2\pi l/M)$ with generic rotation angle $\theta=2\pi/M$ with $M\in \Z$. We will see that the partial rotation with the large rotation angle $M^2\ll L/\xi$ behaves quite differently from the small rotation angle $L/\xi\ll M^2$. Unlike the case with the large rotation angle, $\mathcal{T}^b(2\pi/M,2\pi l/M)$ with the small rotation angle gives a non-universal value that depends on the correlation length of the system. At the same time, it also depends on the universal responses such as Hall conductivity.

\begin{itemize}
    \item $M^2\ll L/\xi$. 
This case corresponds to the analysis in Sec.~\ref{subsec:cftpermutation}, and the partial rotation with $l=0$ is given by Eq.~\eqref{eq:partialrot_boson_U(1)_permute}. When the rotation symmetry does not permute anyons, the expression simplifies as Eq.~\eqref{eq:nonpermuteU1_app}.

\item $L/\xi\ll M^2$. Let us assume that the rotation symmetry does not permute anyons.

When $M$ satisfies $L/\xi \ll M^2$, one cannot follow the logic in Sec.~\ref{subsec:cftpermutation} since the approximation of the character \eqref{eq:character_approx}
is no longer valid. Instead of the modular $ST^MS$ transformation to the CFT character performed in App.~\ref{app:cft}, let us do $S$ transformation with generic nonzero $l$
\small
\begin{align}
\begin{split}
        &\chi_1\left(\frac{i\xi}{L}-\frac{1}{M};(0,0),(1,\frac{2\pi l}{M})\right) \\
        &= \sum_a S_{1a} \chi_a\left(-\frac{1}{\frac{i\xi}{L}-\frac{1}{M}};(1,\frac{2\pi l}{M}),(0,0)\right) 
        \end{split}
\end{align}
\normalsize
where $(0,0), (1,\frac{2\pi l}{M}))\in \Z_M\times U(1)$ represents the twist along each direction of the torus.
One can then use the alternative approximation
\small
\begin{align}
\begin{split}
    &\sum_a S_{1a}\chi_a\left(-\frac{1}{\frac{i\xi}{L}-\frac{1}{M}};(1,\frac{2\pi l}{M}),(0,0)\right) \\
    &\approx \frac{1}{\mathcal{D}} \exp\left(-{2\pi i(h_{0_{(1,\frac{2\pi l}{M})}}-\frac{c_-}{24})\frac{1}{\frac{i\xi}{L}-\frac{1}{M}}}  \right)
    \end{split}
\end{align}
\normalsize
where $h_{0_{(1,\frac{2\pi l}{M})}}$ is the lowest energy of the $(1,{\frac{2\pi l}{M}})$-twisted sector.
When the CFT is chiral, the spin of the $\mu\in U(1)$ vortex $h_\mu$ is given by~\cite{schwimmer1987comments, fan2022}
\begin{align}
    2\pi h_{0_{\mu}} = \frac{k}{2}\left(\frac{\mu}{2\pi}\right)^2, 
\end{align}
where $k$ is the level of the holomorphic $U(1)$ current algebra. It is related to the electric Hall conductivity as $\sigma_H=k/2\pi$.
Using the above relation, one can write the $\Z_M\times U(1)$ vortex as
\begin{align}
    h_{0_{(1,\frac{2\pi l}{M})}} = \frac{\sigma_H l^2}{2M^2} + \frac{\mathscr{S}l}{M^2} + \frac{\ell_s}{2M^2},
\end{align}
where $\sigma_H,\mathscr{S},\ell_s$ are coefficients of the response action
\begin{align}
    -\frac{\sigma_H}{4\pi}Ad  A - \frac{\mathscr{S}}{2\pi}Ad  \omega - \frac{\ell_s}{4\pi}\omega d\omega
\end{align}
where $\omega$ is a $SO(2)=U(1)$ gauge field for continuous spatial rotation symmetry.
The partial rotation is then given by
\begin{widetext}
\begin{align}
\begin{split}
    \mathcal{T}_1\left(\frac{2\pi}{M},\frac{2\pi l}{M}\right) =&e^{-\frac{2\pi i}{24M}c_-}\exp\left(-{2\pi i \left(\frac{\sigma_H l^2}{2M^2} + \frac{\mathscr{S}l}{M^2} + \frac{\ell_s}{2M^2}-\frac{c_-}{24}\right)\frac{1}{\frac{i\xi}{L}-\frac{1}{M}}}-\frac{2\pi L}{\xi}\frac{c_-}{24}\right) \\
    =&e^{-\frac{2\pi i}{24M}c_-}\exp\left({2\pi i \left(\frac{\sigma_H l^2}{2M^2} + \frac{\mathscr{S}l}{M^2} + \frac{\ell_s}{2M^2}-\frac{c_-}{24}\right)\frac{\frac{1}{M}}{\frac{\xi^2}{L^2} + \frac{1}{M^2}}}\right) \\
    &\times \exp\left( {-2\pi  \left(\frac{\sigma_H l^2}{2M^2} + \frac{\mathscr{S}l}{M^2} + \frac{\ell_s}{2M^2}-\frac{c_-}{24}\right)\frac{\frac{\xi}{L}}{\frac{\xi^2}{L^2} + \frac{1}{M^2}}}-\frac{2\pi L}{\xi}\frac{c_-}{24}\right) \\
\end{split}
\end{align}
\end{widetext}
\end{itemize}

\section{More on the classification of symmetry-enriched topological phases}\label{app:Realspace}
\subsection{Equivalences on $([s_\alpha], [s_\beta], [s_\gamma])$}

\begin{figure}
    \centering
    \includegraphics[width=0.49\textwidth]{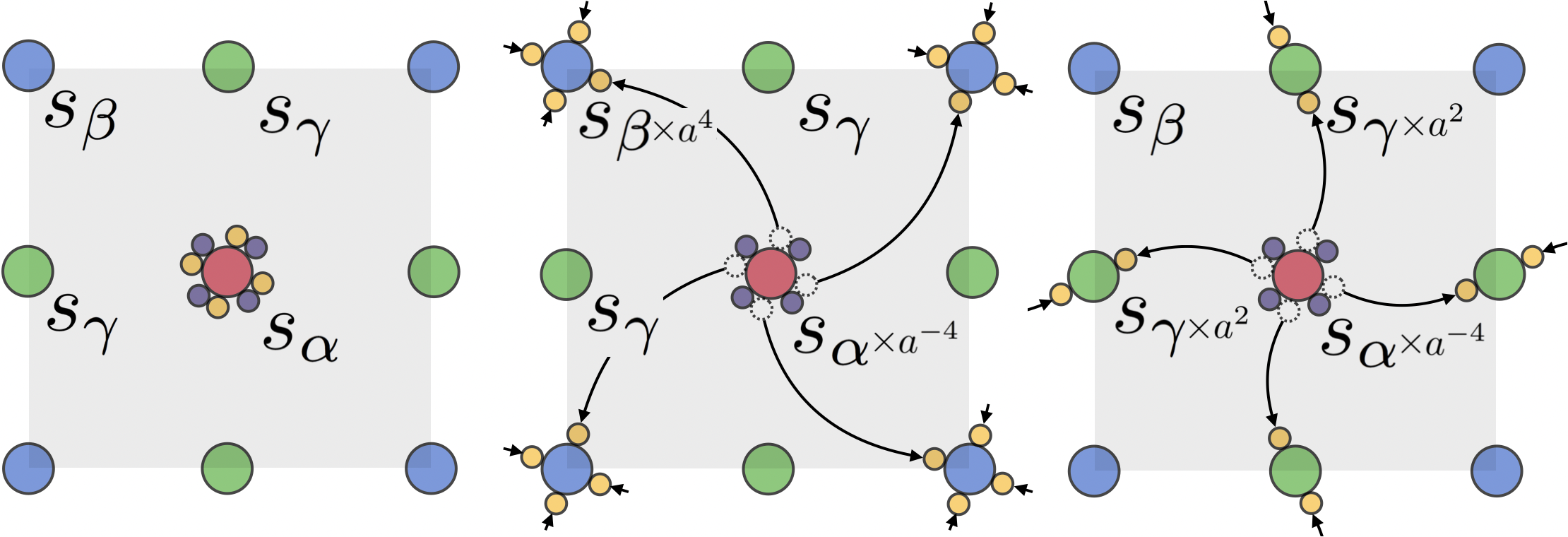}
    \caption{Equivalences in the real-space symmetry fractionalization data. \textbf{Left. } We place abelian anyons $s_\OO$ at the high symmetry points and create four pairs of anyons $(a,a^{-1})$ (orange and purple bubbles respectively) at $\alpha$. \textbf{Middle.} We symmetrically move only the $a$ anyons to $\beta$,  implementing Eq.~\eqref{eq:equiv-1}. \textbf{Right.}
    We symmetrically move only the $a$ anyons to 
    $\gamma$, implementing Eq.~\eqref{eq:equiv-2}.}
    \label{fig:anyon-movement}
\end{figure}

As stated in the main text, for a general topological order on the square lattice, the assignment $([s_\alpha], [s_\beta], [s_\gamma])$ can be adiabatically modified. For example, we can create four pairs of the Abelian anyons $a,a^{-1}$ at $\alpha$ and move the $a$ anyons symmetrically to $\beta$ or $\gamma$ (See Fig.~\ref{fig:anyon-movement}). This changes the assignment, giving us the equivalences
\begin{align}
    ([s_\alpha], [s_\beta], [s_\gamma]) &\simeq ([s_{\alpha} \times a^{-4}],[s_{\beta} \times a^4],[s_{\gamma}]) \label{eq:equiv-1} \\
    ([s_\alpha], [s_\beta], [s_\gamma]) &\simeq ([s_{\alpha} \times a^{-4}],[s_{\beta}] , [s_{\gamma}\times a^2] ).\label{eq:equiv-2}
\end{align}
The full symmetry fractionalization classification is given by the number of possible triples $([s_\alpha], [s_\beta], [s_\gamma])$, modulo the above equivalences. Since $a^2$ is trivial in $\U_{\pm 2}$, this does not affect our case. But more generally, if the Abelian anyons form a group $\mathcal{A}$ under fusion, the final classification from this procedure turns out to be $\mathcal{A} \times (\mathcal{A}/4\mathcal{A}) \times (\mathcal{A}/2\mathcal{A})$, where $n\mathcal{A}$ denotes the group of Abelian anyons $a^n$, where $a \in \mathcal{A}$. By comparing with Ref.~\cite{manjunath2021cgt}, we see that this equals the cohomology group $\mathcal{H}^2(\text{p4},\mathcal{A})$, which is the expected mathematical classification of symmetry fractionalization from the G-crossed braided tensor category approach.

Note that the anyon per unit cell $m$ is invariant under the equivalences on $s_{\alpha},s_{\beta},s_{\gamma}$ in general. Therefore the $\mathcal{A}$ factor in the classification can be understood as the possible choices of $m$. The $\mathcal{A}/4\mathcal{A}$ factor corresponds to a choice of $s_{\OO}$ for $\OO = \alpha$ or $\beta$ up to the equivalences above, while the remaining $\mathcal{A}/2\mathcal{A}$ factor corresponds to a choice of either $s_{\gamma}$, or the torsion anyon $t$, which we define below.

\subsection{Discrete torsion vector in terms of real space construction}\label{app:torsion}

Refs.~\cite{manjunath2021cgt,manjunath2020FQH} introduced a discrete torsion vector $\vec{t} = (t_{x}, t_{y})$, which is a pair of Abelian anyons that partially characterizes the crystalline symmetry fractionalization. Here we give an intuitive understanding of $\vec{t}_{\OO}$ using the notion of anyonic polarization. 

According Refs.~\cite{manjunath2021cgt,manjunath2020FQH} to, on the square lattice, $\vec{t}_{\OO}$ assigns an anyon $t_{\OO,x}^a t_{\OO,y}^b$ to a region with Burgers vector $(a+b, a-b)$, up to an equivalence relation. Braiding another anyon $c$ around a region containing such a defect then gives a phase given by the mutual braiding between $c$ and $t_{\OO,x}^a t_{\OO,y}^b$. In particular, a dislocation with Burgers vector $(2,0)$ is assigned an anyon $t_{\OO} := t_{\OO,x} t_{\OO,y}$. In the present example, $t_{\OO}$ is an invariant under the equivalence relations of \cite{manjunath2021cgt,manjunath2020FQH} and completely characterizes the inequivalent choices of $\vec{t}_{\OO}$. We can understand this heuristically in the real-space construction as follows. 
Observe that $s_\alpha$, $s_\beta$, $s_\gamma$ formally defines a topological charge (anyonic) polarization $\vec{p}_o$. $\vec{p}_\beta$ is formally defined as $\vec{p}_{\beta} = ([s_\gamma + s_\alpha]^{1/2}, [s_\gamma + s_\alpha]^{1/2})$, and the topological charge polarization with respect to $\alpha$ is $\vec{p}_\alpha = ([s_\gamma +  s_\beta]^{1/2}, [s_\gamma + s_\beta]^{1/2})$. The fundamental property of polarization is that a dislocation with Burgers vector $\vec{b}$ is assigned a charge $\vec{p}_\beta \cdot \vec{b}$~\cite{zhang2022pol}. Thus for $\vec{b} = (2,0)$, the region is assigned the anyon $[s_\alpha + s_\gamma]$. This motivates us to define $t_{\alpha} = s_{\alpha} s_{\gamma}$, thus explaining the relationship between $t_o$ and $s_o$. 

The above analysis was heuristic, as $[s]^{1/2}$ is not a well-defined object in the mathematical theory. A more technical explanation that is in line with the G-crossed BTC description is given below. Consider a general space group operation ${\bf g} = (\vec{r},{\bf h})$, where ${\bf h}$ is an \textit{elementary} rotation by the angle $2\pi/M_{\OO}$ about some origin $\OO$, and ${\bf r}$ is a lattice translation. We have $M_{\OO}=4$ for $\OO = \alpha,\beta$ and $M_{\OO}=2$ for $\OO=\gamma$. Note that 
\begin{equation}
    {\bf g} = (\vec{r},0) \times (\vec{0},{\bf h}) =  (\vec{0},{\bf h}) \times (^{\bf \bar{h}}\vec{r},0) 
\end{equation}
where ${^{\bf \bar{h}}}\vec{r}$ is the vector ${\bf r}$ rotated by ${\bf h}^{-1}$. This implies that we can obtain a ${\bf g}$ defect in two ways: by fusing an ${\bf h}$-disclination at $\OO$ to a dislocation with Burgers vector $\vec{r}$ (we will call this an $\vec{r}$-dislocation) where the dislocation is to the left, or fusing the ${\bf h}$-defect to an ${^{\bf \bar{h}}}\vec{r}$-dislocation where the dislocation is to the right. The main point is that when there is nontrivial symmetry fractionalization, the two fusion processes will differ by the anyon $\vec{t}_{\OO} \cdot \vec{r} := t_{\OO,x}^{r_x} \times t_{\OO,y}^{r_y}$. Equivalently, $\vec{t}_{\OO} \cdot \vec{r}$ is the residual anyon left behind at an ${\bf h}$-disclination, upon dragging an ${\bf r}$-dislocation through it from left to right. 

Let us denote a reference ${\bf g}$ defect by $0_{\bf g}$. The subscript denotes the group element associated to the defect, and for Abelian topological order without anyon-permuting symmetries, the number of ${\bf g}$-defects is given by the number of Abelian anyons, for any ${\bf g}$ \cite{barkeshli2019}. When ${\bf g}$ generates a discrete subgroup, we can pick any of the above defects as our $0_{\bf g}$. The other ${\bf g}$-defects are related to it as $a_{\bf g} := a \times 0_{\bf g}$ where $a$ is an Abelian anyon. Now from the above discussion it follows that
\begin{equation}
    0_{\vec{r}} \times 0_{\bf h}  = \vec{t}_{\OO} \cdot \vec{r} \times 0_{\bf h} \times 0_{{^{\bf \bar{h}}}\vec{r}}.
\end{equation}

For the square lattice, we will now derive the relationship
\begin{equation}\label{eq:t-realspace}
    t_{\alpha(\beta)} = s_{\alpha(\beta)} s_{\gamma} \mod a^2, a \in \mathcal{A}.
\end{equation}
The derivation is as follows. Let $\OO = \alpha$, and ${\bf h}$ be a $\pi/2$ rotation about $\alpha$. We know that $s_{\alpha}$ is the anyon induced by fusing four ${\bf h}$ disclinations:
\begin{equation}
    s_{\alpha} = 0_{\bf h}^4 \mod a^4.
\end{equation} 
But note that $(\hat{x},0) \times {\bf h}^2$ is actually a $\pi$ rotation about $\gamma$. Therefore $s_{\gamma}$ is induced by fusing two $(\hat{x},0) \times {\bf h}^2$ defects: 
\begin{align}
   s_{\gamma} = (0_{\vec{x}} \times 0_{\bf h}^2)^2 \mod a^2.  
\end{align}
But we can drag the $\vec{x}$ dislocation defects through the ${\bf h}$-defects and obtain a relation between  
 $s_{\gamma}$ and $s_{\alpha}$:
 \begin{align}
   s_{\gamma} &=  0_{\vec{x}} \times 0_{\bf h}^2 \times 0_{\vec{x}} \times 0_{\bf h}^2 \nonumber \\
   &= t_{\alpha,x} \times 0_{\bf h} \times 
 0_{\vec{y}} \times 0_{\bf h} \times 0_{\vec{x}} \times 0_{\bf h}^2 \nonumber \\
   &= t_{\alpha,x} t_{\alpha,y} \times 0_{\bf h}^2 \times 0_{-\vec{x}} \times0_{\vec{x}} \times 0_{\bf h}^2 \nonumber \\
   &= t_{\alpha,x} t_{\alpha,y}\times  0_{\bf h}^4 \nonumber \\ &= t_{\alpha,x} t_{\alpha,y} \times s_{\alpha} =: t_{\alpha} \times s_{\alpha} \mod a^2.
 \end{align}
 Here we chose a gauge in which we can trivially fuse defects which are not important for symmetry fractionalization, such as $0_{\vec{x}}$ and $0_{-\vec{x}}$. This does not affect the final result. Moreover, since $t_{\alpha}$ is defined mod $a^2$, we can equivalently write the last line as $t_{\alpha} \times s_{\alpha}^{-1}$, for any topological order in which the symmetry does not permute anyons. This gives the claimed result.

\subsection{Relabeling and modular reduction}\label{app:relabel}

Let ${\bf h}$ be the group element associated to $\hat{C}_{4,l}$. Note that the invariant $\mathcal{I}_{M,\theta=2\pi l/M}$ in Eq.~\eqref{eq:IMtheta} computes the $\hat{C}_{4,l}$ charge associated to a reference $\hat{C}_{4,l}$ defect which we denote by $0_{\bf h}$. But as we emphasized in the previous section, there is no canonical choice of this reference defect; we can always redefine it by fusing an Abelian anyon $a$, so that $0_{\bf h} \rightarrow 0_{\bf h}\times a =: a_{\bf h}$. This will change the value of $\mathcal{I}_{M,\theta=2\pi l/M}$ by various braiding phases associated to $a$. Ref.~\cite{manjunath2020FQH} showed that the invariants before and after relabelling the defect are related as follows:
\begin{equation}\label{eq:bSET-relabel-general}
    \frac{\mathcal{I}_{M,\theta=2\pi l/M}(a_{\bf h})}{\mathcal{I}_{M,\theta=2\pi l/M}(0_{\bf h})} = M_{s^*,a} \times \theta_a^M
\end{equation}
where $s^* = 0_{\bf h}^M$ is an anyon that we determine below from the symmetry fractionalization data (charge/spin vectors). Since there is no canonical choice of an elementary $C_{4,l}$ defect, two systems in which the partial rotation phases differ by this amount should be treated as equivalent under relabellings.
 
We denote by $s^*$ the anyon induced upon inserting $M$ elementary $C_{4,l}$ defects, that is, disclinations with total angle $2\pi$ together with $2\pi l$ flux of $U(1)$. This implies $s^* = s \times v^l$. For $\U_2$ topological order, we further have $s^* = a^{s + v l}$ where $a$ is a semion. The right-hand side of Eq.~\eqref{eq:bSET-relabel-general} now reduces to
\begin{equation}
    M_{s^*,a}\theta_a^M = \begin{cases}
      (-1)^{s+vl} \times i^4= (-1)^{s + v l}  & \OO = \alpha,\beta \\
       (-1)^{1+s + v l} & \OO = \gamma.
    \end{cases}
\end{equation}
After modding out by these quantities, the partial rotation invariant is reduced modulo $\gcd(4,2(s+vl))$ for $\OO = \alpha,\beta$ (see Eq.~\eqref{eq:Theta-_vs_k3} of the main text), and mod $\gcd(2,1+s + vl)$ for $\OO = \gamma$ (see Eq.~\eqref{eq:Theta-gamma_vs_k}).

\vfill

\end{document}